\DocumentMetadata{ 
	pdfstandard = a-2b,
	pdfversion  = 1.7,
	lang		= en-US,
}

\documentclass[openany]{mitthesis} 
\usepackage{listings}

\usepackage[version=4]{mhchem}

\usepackage{lipsum}
\IfPackageAtLeastTF{lipsum}{2021/09/20}{\setlipsum{auto-lang=false}}{}

\usepackage[style=ieee,maxbibnames=10,sorting=none]{biblatex}
	\DefineBibliographyStrings{english}{url= \textsc{url} ,  }

\usepackage{booktabs}
\usepackage{array}

\usepackage{amsfonts}
\usepackage{tikz}
\usepackage{amsmath}
\usepackage{algorithm}
\usepackage{algpseudocode}
\usepackage[acronym,nomain,nonumberlist,automake]{glossaries}
\makeglossaries
\usepackage{typed-checklist}
\usepackage{enumitem}
\usepackage{longtable}

\usepackage{hyperref}
\usepackage{cleveref}
\usepackage[colorinlistoftodos]{todonotes}
\usepackage{multirow}
\usepackage{graphicx}
\usepackage[stable]{footmisc}
\usepackage{svg}

\Crefname{figure}{Fig.}{Figs.}
\crefname{figure}{fig.}{figs.}

\DeclareMathAlphabet{\mathcal}{OMS}{cmsy}{m}{n}
\SetMathAlphabet{\mathcal}{bold}{OMS}{cmsy}{b}{n}

\usetikzlibrary{shapes.geometric, arrows, positioning}
\tikzstyle{startstop} = [rectangle, rounded corners, minimum width=3cm, minimum height=1cm,text centered, draw=black]
\tikzstyle{io} = [trapezium, trapezium left angle=70, trapezium right angle=110, minimum width=3cm, minimum height=1cm, text centered, draw=black]
\tikzstyle{process} = [rectangle, minimum width=3cm, minimum height=1cm, text centered, draw=black]
\tikzstyle{decision} = [diamond, minimum width=3cm, minimum height=1cm, text centered, draw=black]
\tikzstyle{arrow} = [thick,->,>=stealth]

\AtBeginEnvironment{appendices}{\crefalias{chapter}{appendix}}

\hypersetup{%
	pdfsubject={Template for writing MIT theses with the mitthesis class},
	pdfkeywords={Carnegie Mellon University, CMU},
	pdfurl={},
	pdfcontactemail={apillay@cmu.edu},
	pdfauthortitle={},
}

\begin{document}


\title{A Neural Score Follower for Computer Accompaniment of Polyphonic Musical Instruments}



\Author{Ashwin Pillay}{School of Music}

\Degree{Master of Science in Music and Technology}{School of Music}

\Supervisor{Dr. Richard M. Stern}{Professor of Electrical and Computer Engineering}
\Supervisor{Dr. Roger B. Dannenberg}{Emeritus Professor of Computer Science, Art and Music}


\DegreeDate{May}{2024}

\ThesisDate{May 16, 2024}

%
%
\CClicense{CC BY-NC-ND 4.0}{https://creativecommons.org/licenses/by-nc-nd/4.0/}
%


%
%
%
%
%
%
%
%
%

%
%


\maketitle




\begin{abstract}
	Real-time computer-based accompaniment for human musical performances entails three critical tasks: identifying what the performer is playing, locating their position within the score, and synchronously playing the accompanying parts. Among these, the second task (score following) has been addressed through methods such as dynamic programming on string sequences, Hidden Markov Models (HMMs), and Online Time Warping (OLTW). Yet, the remarkably successful techniques of Deep Learning (DL) have not been directly applied to this problem.

Therefore, we introduce \textbf{HeurMiT}, a novel DL-based score-following framework, utilizing a neural architecture designed to learn compressed latent representations that enables precise performer tracking despite deviations from the score. Parallelly, we implement a real-time MIDI data augmentation toolkit, aimed at enhancing the robustness of these learned representations. Additionally, we integrate the overall system with simple heuristic rules to create a comprehensive framework that can interface seamlessly with existing transcription and accompaniment technologies.

However, thorough experimentation reveals that despite its impressive computational efficiency, HeurMiT's underlying limitations prevent it from being practical in real-world score following scenarios. Consequently, we present our work as an introductory exploration into the world of DL-based score followers, while highlighting some promising avenues to encourage future research towards robust, state-of-the-art neural score following systems.
\end{abstract}



\tableofcontents
\listoffigures
\listoftables

\newacronym{hmm}{HMM}{Hidden Markov Models}
\newacronym{oltw}{OLTW}{Online Time Warping}
\newacronym{dtw}{DTW}{Dynamic Time Warping}
\newacronym{stft}{STFT}{Short-time Fourier Transform}
\newacronym{mirex}{MIREX}{Music Information Retrieval Evaluation eXchange}
\newacronym{nmf}{NMF}{Non-negative Matrix Factorization}
\newacronym{cqt}{CQT}{Constant-Q Transform}
\newacronym{dl}{DL}{Deep Learning}
\newacronym{mdtk}{MDTK}{MIDI Degradation Toolkit}
\newacronym{dnn}{DNN}{Deep Neural Networks}
\newacronym{rl}{RL}{Reinforcement Learning}
\newacronym{crf}{CRF}{Conditional Random Fields}
\newacronym{drl}{DRL}{Deep Reinforcement Learning}
\newacronym{daw}{DAW}{Digital Audio Workstation}
\newacronym{ae}{AE}{Alignment Error}
\newacronym{dsp}{DSP}{Digital Signal Processing}
\newacronym{cnn}{CNN}{Convolutional Neural Network}
\newacronym{gdl}{GDL}{Geometric Deep Learning}
\newacronym{amt}{AMT}{Automatic Music Transcription}
\newacronym{cli}{CLI}{command-line interface}
\newacronym{gil}{GIL}{Global Interpreter Lock}
\newacronym{osc}{OSC}{Open Sound Control}
\newacronym{udp}{UDP}{User Datagram Protocol}
\newacronym{api}{API}{Application Programming Interface}
\newacronym{sd}{SD}{standard deviation}
\newacronym{relu}{ReLU}{Rectified Linear Unit}
\newacronym{nsgt}{NSGT}{Non Stationary Gabor Transform}
\newacronym{lstm}{LSTM}{Long Short-Term Memory}

\glsaddall
\printglossary[type=\acronymtype]



\chapter{Introduction}
\label{ch:introduction}

Music has always been a dynamic art form that readily embraces technological innovations to expand its expressivity and accessibility. The advent of modern computing has transformed music production from a resource-intensive process, which traditionally required significant manpower, expertise, and financial investment, into an activity that individuals can pursue effectively within the comfort of their own rooms.

However, computers are yet to be widely applied in live collaborations with human musicians. Enabling computers to act as accompanists to human performers introduces a new dimension to live performances. They could simultaneously control multiple instruments or even non-musical elements such as lighting and stage visuals. Crucially, we strongly intend these collaborative performances to be human-led; \emph{i.e.}, while both the human performer and the computer have access to the score, the human dictates the timing and tempo of the performance. The computer must listen to the human's performance and play the rest of the score, intelligently navigating through any mistakes or improvisations made by the performer. This setup allows human musicians to maintain their creative expression while benefiting from the quick, error-free response, and multitasking capabilities of computer systems.

The problem of computer-based accompaniment has been the subject of previous research. In 1984, Dannenberg \cite{Dannenberg1984AnOA} identified the main challenges in this domain, segmenting them into three sub-problems: identifying what a performer is playing, locating their position within the score (termed \textit{score following}), and playing the remaining score components to accompany the performer with minimal latency.  With respect to score following, various strategies ranging from dynamic programming to \gls{hmm}, have been employed but the potential of \gls{dl} remains to be fully leveraged. Recent advances in \gls{dl} have revolutionized fields such as text understanding \cite{Touvron2023Llama2O}, speech recognition \cite{Radford2022RobustSR}, and image generation \cite{Ramesh2021ZeroShotTG}, suggesting promising new directions for enhancing score following systems.

In this work, we develop \textbf{HeurMiT}, a score following system that comprises the following:
\begin{enumerate}
    \item \textbf{Tyke}, a compact \gls{dnn}-based architecture trained using a template-matching \gls{dl} paradigm. Tyke aims to learn compressed latent feature representations of the performance and the score while being robust against any performer-induced deviations.
    \item A set of common-sense heuristic rules that critique and safeguard Tyke's predictions, independently of any performance-specific nuances.
\end{enumerate}

For training Tyke models to learn features robust to the performer's random deviations from the score, we have also developed an on-the-fly MIDI data augmentation library called \textbf{MIDIOgre}. This tool can synthetically generate a wide variety of commonly observed performance imperfections, which we incorporate into Tyke's training paradigm.

The chief benefits of HeurMiT include its \(\mathcal{O}(1)\) performance.
Additionally, its design enables easy integration with existing solutions addressing the other sub-problems outlined by Dannenberg. However, a thorough evaluation of its score following ability suggests that HeurMiT is \textbf{not ready} for real-world applications. In a best-case scenario, HeurMiT's prediction errors are only slightly better than our baseline system (Flippy), while it loses track of the performance more frequently on average. Moreover, when the performance tempo significantly differs from the score, HeurMiT fails to follow the performance comprehensively. Additionally, the implemented set of heuristics continues to focus on performance nuances, necessitating optimizations per performance for optimal results. In light of these limitations, we discuss several directions for future research that either seek to improve the current template-matching paradigm or identify alternative \gls{dl}-based approaches that are inherently tempo-insensitive.

We view our work as an introductory exploration into the capabilities of \gls{dl}-based neural score following systems. By documenting our approach and acknowledging both the strengths and limitations of our efforts, we aspire to encourage future research aimed at identifying much more robust and efficient methods for tracking performances across a diverse array of instruments, genres, and styles using \gls{dnn}s. We believe continued efforts in this space will eventually enable effective real-time musical collaboration between humans and computers.

The subsequent sections of this work are structured as follows: \Cref{ch:Literature.Review} discusses existing literature related to score following, computer accompaniment, and recent advancements in \gls{dl}. \Cref{ch:Proposed.System} introduces a formal definition of our \gls{dnn}-based score following problem, detailing the neural architectures and training paradigms employed, along with auxiliary components addressing the first and third sub-problems. \Cref{ch:Evaluation} outlines the metrics and experiments designed to evaluate our system comprehensively against existing works and set a benchmark for future comparisons. \Cref{ch:results} provides a detailed analysis of the performances of our system, highlighting the pros and cons of our approach. \Cref{ch:Future.Directions} presents a set of future research directions for neural score following systems. Finally, \cref{ch:Conclusion} concludes our research.
\chapter{Literature Review}
\label{ch:Literature.Review}

\glsresetall

\section{Foundational Work based on Dynamic Programming}
\label{secn:early.work}

The concept of \textit{score following} and its extension to real-time applications, termed \textit{computer accompaniment}, was independently introduced by Dannenberg \cite{Dannenberg1984AnOA} and Vercoe \cite{Vercoe1984TheSP} in 1984. This pioneering work laid the foundation for pitch tracking-based score following systems, where incoming performance audio is analyzed through pitch detection—and, in certain cases, supported by additional data sources like fingering information and optical cues—to generate a sequence of musical notes.

Dannenberg's approach conceptualizes score following as a symbol-matching challenge, converting score and performance events into string sequences. The goal is to identify the least-cost alignment path between these sequences using dynamic programming techniques. Originally designed for monophonic instruments such as the trumpet, this method was later expanded to accommodate polyphonic performances \cite{Bloch1985RealTimeCA}, enhance resilience to performance interruptions through multiple matchers running in parallel \cite{Dannenberg1988NewTF}, and track complex musical ornaments, including trills and glissandi \cite{Dannenberg1988NewTF}. Combining ideas from Dannenberg \cite{Dannenberg1984AnOA} and Vercoe \cite{Vercoe1984TheSP}, Vercoe and Puckette proposed an alternative approach based on least-cost matching \cite{Vercoe1985SyntheticRT}. This methodology has since evolved to include considerations for fixed-length note segments \cite{Baird1993ArtificialIA}, the ability to match new notes with previously overlooked ones \cite{Puckette1992ScoreFI}, and the integration of pitch detection with note onset and tempo tracking \cite{Vantomme1995ScoreFB}.

\section{Statistical Approaches}
\label{secn:statistical.approaches}

A fundamental objective for score following systems is to remain resilient against the variability and inaccuracies inherent in live human performances, especially regarding timing. Early systems employed heuristic rules informed by musical theory and general common sense. However, subsequent generations have adopted probabilistic statistical models to enhance accuracy and flexibility. For instance, in 1997, Grubb and Dannenberg \cite{Grubb1997ASM} introduced a method that uses a sliding window to generate a continuous probability density function over possible score positions for vocal performance tracking. Pardo and Birmingham \cite{Pardo2002ImprovedSF} refined the least-cost matching strategy with a probabilistic model that accounts for audio-to-symbol transcription errors and introduces penalties for omitting significant portions of the score.

\subsection{\gls{hmm}}
\label{subsecn:hmm}

Among the statistical models employed in score following, \gls{hmm}s have become particularly prominent. Raphael's 1999 \cite{Raphael1999AutomaticSO} proposal of an \gls{hmm}-based score follower that operates directly on audio spectral features—bypassing the need for pitch-to-symbol conversion—marked a significant advancement. This model was further developed to account for performer errors \cite{Orio2001ScoreFU}, support polyphonic music \cite{Schwarz2004RobustPM}, and handle impromptu skips and repeats in the performance \cite{Pardo2005ModelingFF}. Subsequently, several performance improvement extensions have also ensued \cite{Cont2004ImprovementOO,Cont2006RealtimeAT,Cont2008AntescofoAS,Cuvillier2014CoherentTM,Montecchio2008AutomaticAO, Nakamura2014OuterProductHM,Sagayama2014AutomaticMA,Hori2019PianoPE}, with works being as recent as from 2023 \cite{Chacn2023TheAC}.

\section{\gls{oltw}-based Approaches}
\label{subsecn:dtw}

Following the development of \gls{oltw} \cite{Dixon2005LIVETO}, a variant of \gls{dtw} characterized by linear space and time complexity, there was an advent of a new subclass of score followers. These facilitate incremental alignment of real-time performance audio with a synthesized version of the score using their \gls{stft} analyses. In this direction, studies have expanded upon the work by
Dixon and Widmer \cite{Dixon2005MATCHAM} to be run online, while aiming to minimize the discrepancies between offline and online \gls{dtw} alignments \cite{Arzt2008AutomaticPT,RodrguezSerrano2016TempoDA}. Notably, many systems benchmarked by the \gls{mirex} \textit{Real-time Audio to Score Alignment (a.k.a Score Following)} task utilize \gls{oltw} in conjunction with advanced feature extraction methods, such as chromagrams \cite{Suzuki2010REALTIMEAT} and \gls{nmf} bases \cite{CarabiasOrti2015AnAT}. A recent study by Lee \cite{Lee2022MusicalSF} also demonstrates the effectiveness of combining \gls{oltw} with \gls{cqt}.

\section{Breakthroughs in \gls{dl} and Intersections with Computer Accompaniment}
\label{subsecn:dl.and.computer.accompaniment}
In parallel with the progression of score following methodologies, \gls{dl} \cite{LeCun2015DeepL} has risen to prominence through leveraging neural networks as universal function approximators \cite{Hornik1989MultilayerFN}. This approach has become the go-to solution for a plethora of challenges across unimodal and multimodal applications in text, vision, video, and audio domains, among others
\cite{Touvron2023Llama2O,Kim2023ConsistencyTM,Rouard2022HybridTF,Gupta2023PhotorealisticVG}. A pivotal aim of \gls{dl} models is the extraction of optimal features. This is usually achieved by training neural networks on extensive sets of input-output examples to derive a condensed representation within a learned latent space. Such extracted features are now integral to cutting-edge solutions tackling complex problems such as vision-language understanding \cite{Lu2024DeepSeekVLTR}, speech processing \cite{Liu2021DiffSingerSV}, and code generation \cite{Li2023StarCoderMT}. Data augmentation stands as a critical strategy for deriving efficient latent representations, wherein training data is enriched through artificial modifications of minor information contributors, thereby rendering models invariant to their random real-world variabilities, including noise. In the vision domain, data augmentation strategies range from simple techniques like cropping, rotations, and color space transformations \cite{Shorten2019ASO} to more complex methods such as CutMix \cite{Yun2019CutMixRS}. For audio and speech, basic techniques include random pitch shifting and time stretching \cite{Wei2020ACO}, with advanced methods like SpecAugment \cite{Park2019SpecAugmentAS} also proving beneficial. In contrast, the exploration of effective augmentation techniques for symbolic music (e.g., MIDI, MusicXML, piano rolls) remains nascent despite their increasing utility \cite{Thickstun2023AnticipatoryMT,Zeng2021MusicBERTSM}. However, the \gls{mdtk} \cite{Mcleod2020TheMD} initiates an exploration into effective MIDI augmentations like random note number shifts, velocity shifts, and timing adjustments.

Recent advancements in computer accompaniment have also leveraged \gls{dnn}s. Jiang et al. introduced RL-Duet \cite{Jiang2020RLDuetOM}, a \gls{rl} based generation agent that devises a policy for musical note generation conditioned on prior human and machine inputs. This showcases \gls{rl} agents' capability in maintaining long-term tempo coherence through maximizing discounted rewards. Wang et al. presented SongDriver \cite{Wang2022SongDriverRM}, a real-time music generator employing a Transformer \cite{Vaswani2017AttentionIA} and a \gls{crf} model for anticipative melody accompaniment with minimal latency. These studies primarily focus on \textit{score-free} accompaniment, posing a likelihood for it to significantly diverge from the score. A conceivable improvement involves anchoring these models on partial score information via a Music ControlNet \cite{Wu2023MusicCM} though more direct score-based training approaches might offer optimized solutions. In 2023, Peter \cite{Peter2023OnlineSM} proposed an \gls{rl} agent, utilizing an attention-based neural network \cite{Bahdanau2014NeuralMT}, trained via Offline \gls{drl} to align new performance data with the current score window and the most recent performance data. For real-time score following, this method integrates a tempo extractor and heuristic rules to bolster overall robustness.

\section{Evaluating Existing Research}
\label{section:evaluating.existing.research}

This section builds on our analysis of existing methodologies in score following and computer accompaniment by outlining the desired characteristics within an optimal score follower. We examine how these characteristics have been incorporated into existing research and identify how they can be enhanced. While we aim to integrate all the features outlined in this section within our work, we also intend for them to serve as general benchmarks for future developments in the field.

\subsection{Accessibility}
\label{subsection:accessibility}

Performance variability, discounting intentional improvisations, is a function of the performer's expertise. Novices may not adhere closely to the score, posing challenges for score following systems originally conceived for professional performances \cite{Dannenberg1984AnOA,Vercoe1984TheSP,Bloch1985RealTimeCA}. A broader user base, beyond professional musicians, could vastly benefit from using score following systems as an advanced alternative to traditional metronomes. To cater to this demographic, future systems should accommodate varying levels of performance expertise, allowing for substantial deviations as long as the overall piece remains recognizable. Additionally, ease of installation and operation on widely available computing systems is crucial for widespread accessibility.

\subsection{Flexibility}
\label{subsection:Flexibility}

While initial score followers focused on monophonic instruments \cite{Dannenberg1984AnOA,Vercoe1984TheSP,Vercoe1985SyntheticRT,Raphael1999AutomaticSO}, they were subsequently extended for polyphonic music by methods such as grouping notes based on their relative proximities \cite{Bloch1985RealTimeCA}. Additionally, specialized systems have been developed for vocal tracking \cite{Grubb1997ASM}, and some require iterative rehearsals \cite{Vercoe1985SyntheticRT} or score-specific training \cite{Raphael1999AutomaticSO,Pardo2002ImprovedSF,CarabiasOrti2015AnAT}.  Looking forward, an ideal score follower would inherently support polyphonic and vocal performances, minimize dependency on domain-specific heuristics, and efficiently handle a wide array of musical styles and complexities without the need for customized training.

\subsection{Robustness\footnote{In the context of computer accompaniment systems, the term "robustness" has traditionally referred to the system's ability to gracefully recover from drastic jumps in score positions during a performance. In our discussion, however, we define robustness as the system's capacity to tolerate variations in the performance relative to the score.}}
\label{subsection:Robustness}

Many symbol-matching and \gls{oltw}-based algorithms are designed to align performances with scores at the individual note level. However, when performances are uniformly transposed relative to the score, this may pose unique challenges. Several approaches have been developed to account for predictable musical ornaments \cite{Bloch1985RealTimeCA,Dannenberg1988NewTF}, but performers may introduce arbitrary new ornaments. Additionally, significant tempo variations have historically been challenging to accommodate \cite{Schwarz2004RobustPM}. An effective score-following method would need to maintain accuracy amidst these variations. In this context, a holistic template-matching approach that compares collections of performance events with groups of score events might offer a promising solution.

\subsection{Performance Metrics}
\label{subsection:Performance.Metrics}

Throughout the development of score following and computer accompaniment systems, the metrics used to evaluate their performance have significantly evolved. Initially, limitations in computational capabilities may have restricted early studies \cite{Dannenberg1984AnOA,Vercoe1984TheSP,Bloch1985RealTimeCA,Raphael1999AutomaticSO} to qualitative evaluations or minimal quantitative analysis. Subsequent research introduced more standardized measures, such as \gls{ae} \cite{Dixon2005MATCHAM}, assessing the temporal discrepancy between a performance event as identified by the score follower and its actual timing in the score. In 2007, Cont et al. \cite{Cont2007EvaluationOR} expanded the repertoire to include metrics like system latency and precision rate, establishing them as benchmarks within the \gls{mirex} evaluation framework. These metrics have since become a staple in the field \cite{CarabiasOrti2015AnAT,Suzuki2010REALTIMEAT}. Nonetheless, Lee's \cite{Lee2022MusicalSF} critique of the \gls{mirex} evaluation methods highlighted significant limitations, prompting a call for alternative approaches. Despite this, further critical evaluations of the efficacy and applicability of these metrics remain scarce.

\chapter{The Neural Score Following System}
\label{ch:Proposed.System}

Addressing the challenges highlighted in existing score following research, as discussed in \cref{section:evaluating.existing.research}, we formulate a new approach that leverages \gls{dl}. Our strategy involves the conversion of score and performance data into piano roll formats, leveraging \gls{dnn} models to extract robust latent features from them, and applying cross-correlation combined with practical heuristics to jointly compare recent performance data against likely score positions. Furthermore, we explore the \gls{dnn} architecture integral to our framework, elaborating on the supervised learning paradigms utilized for its training. Additionally, we suggest methods to integrate our solution with existing components to enable the accompaniment of polyphonic instruments and vocal performances across a diverse array of musical genres.

\section{Terminologies}
\label{secn:Terminologies}

Aligning with the terminology set forth by Bloch and Dannenberg \cite{Bloch1985RealTimeCA}, we define some key terms for describing our system: the input polyphonic instrument recording is referred to as the \textit{solo}, and the system's output is termed the \textit{accompaniment}. The piece being performed is called the \textit{performance}, with its machine-readable form known as the \textit{score}. Deviations by the performer from the score, which may include tempo variations or unintentional notes, are identified as \textit{imperfections}.

\section{Problem Definition}
\label{secn:problem.defn}

Externally, both the solo and accompaniment manifest as audio signals. However, our score-follower internally interprets piano-roll representations of the solo and the score. Piano rolls are favored for their intrinsic ability to depict notes in both a polyphonic and octave-sensitive manner, while accommodating various musical ornaments such as trills and nachschlags. The conversion from MIDI\footnote{\url{https://midi.org/specs}} and MusicXML to piano rolls is straightforward, with established methods also facilitating their derivation from waveform audios and \gls{stft}s. Mathematically, piano rolls can be represented by two-dimensional vectors, \(p \in \mathbb{R}^{128 \times n}\), where 128 corresponds to the full range of MIDI note numbers\footnote{While most practical systems cover only a subset of this range, we utilize the full set of MIDI note numbers to ensure compatibility with future transcription and accompaniment systems that may accommodate unconventional MIDI note ranges.}, and \(n\) represents the discrete temporal dimension. This structure allows for the efficient application of parallel vector operations available through computational tools like PyTorch \cite{Paszke2019PyTorchAI}.

Building upon the advantages of template-matching, as explored in \cref{subsection:Robustness}, we conceptualize score following as the challenge of locating the most fitting placement of the solo (template) within the larger score (target). Further, aligning with some existing works \cite{Dannenberg1984AnOA,Dixon2005MATCHAM}, we hypothesize that the solo's most probable locations are within a vicinity of its last known position, hence optimizing efficiency by limiting the scope of template-matching to this specific area. This segment of the score, designated as the \textit{context} (\(C \in \mathbb{R}^{128 \times c}\)), alongside the latest collection of notes from the solo, known as the \textit{window} (\(W \in \mathbb{R}^{128 \times w}\)), forms the crux of our analytical framework. Here, \(c\) and \(w\) indicate the durations of score and solo analyzed analyzed for score following, respectively\footnote{While training the \gls{dnn}, we supply batches of \(C\) and \(W\) at once. The batched equivalents are of the form \(C \in \mathbb{R}^{B \times 128 \times c}\) and \(W \in \mathbb{R}^{B \times 128 \times w}\), where \(B\) denotes the batch size used during the training phase.}. Our score-following \gls{dnn} processes these vector representations as its primary inputs.

\subsection{Learning Effective Latent Representations}
\label{subsecn:effective.repr}

When performances are perfect, \gls{dsp} techniques like cross-correlation are effective for template matching. However, our system must be capable of accurate prediction in the presence of the performer's imperfections. Within the piano-roll framework, common imperfections \cite{Mcleod2020TheMD} such as note additions, deletions, delays, and pitch shifts appear as minor vector variations. Transpositions and localized tempo changes can also be represented by correlated pitch shifts and note delays, respectively. We intend the score follower to be insensitive to these variations, while focussing only on the information necessary to reliably locate the window within the score. Additionally, irrespective of the instrument, only a few notes would be simultaneously played at once. This suggests that \(C\) and \(W\) would be sparse, encouraging dimensionality reduction to benefit the efficiency of the algorithm in real-time.

The  concepts described above suggest the development of a neural architecture capable of learning latent representations that are invariant to performance imperfections, yet retain the ability to accurately locate the encoded window (\(W'\)) within the encoded context (\(C'\)). We define these representations as follows:

\begin{equation}
    W' = E_{w}(W | \theta_{w}) \rightarrow \mathbb{R}^{e \times w},
\end{equation}

\begin{equation}
    C' = E_{c}(C | \theta_{c}) \rightarrow \mathbb{R}^{e \times c}.
\end{equation}

Here, \(\theta_{w}\) and \(\theta_{c}\) are the learnable parameters of the neural encoders \(E_{w}\) and \(E_{c}\), respectively, with \(e\) indicating note dimension compression. Since our subsequent cross-correlation operations are performed along the time dimension, we do not compress the inputs along this dimension. 

\(E_{w}\) and \(E_{c}\) can be implemented using 1-D \gls{cnn} layers with a kernel size of \(k\) and a stride of 1, applied after padding \(k // 2\) zeroes on both ends of the input\footnote{"\(//\)" indicates integer division (\emph{i.e.}, we drop the fractional part).}. To describe the underlying operations, consider an \(E_{w}\) architecture consisting of a 1-layer \gls{cnn} such that \(e = 64\) and \(k = 3\), with \(padding=1\) on both ends and \(stride = 1\), resulting in the \gls{cnn} layer having 64 kernels of shape \(128\times3\). This means that for piano-roll window event \(j\), we obtain the element-wise product of weight \(i\) with the previous event (\(W[:, j-1]\)), the current event (\(W[:, j]\)), and the next event (\(W[:, j+1]\)), and sum the result to obtain the latent window element \(W'[i, j]\). Thus, the encoders perform convolution along the time dimension, smoothing out potential imperfections persisting across neighboring piano-roll events for every window position. We also perform a similar operation within \(E_{c}\) for the context, thereby ensuring the downstream cross-correlation occurs within the same latent space.

\subsection{Aligning the Solo with the Performance}
\label{subsecn:Aligning.the.Solo.with.the.Performance}

Assuming the encoded window representation \(W'\) now exactly matches within the context representation \(C'\), score following translates into a classical template-matching task. The goal now is to find the most accurate location of the solo within the score. This is achieved using cross-correlation defined by the following equation:

\begin{equation}
    \label{eqn:xcorr}
    P' = (C' \star W') \rightarrow \mathbb{R}^{1 \times (c+w-1)},
\end{equation}

where \(\star\) denotes the cross-correlation operation. This is implemented by padding \(C'\) on both sides with a zero vector of size \(e \times (w-1)\), and sliding \(W'\) across its discrete time dimension (represented by \(c\)). At each position, we compute the corresponding integer element of \(P'\) by summing the element-wise product of \(W'\) and the overlapping segment of \(C'\). The elements of \(P'\) indicate the likelihood of the window's presence at that context position\footnote{Within the scope of this work, window position specifically refers to the position of its right edge; ie, a window position \(x\) would mean that it occupies score positions from \(x-w\) to \(x\)}. This setup aligns perfectly with the typical \gls{dl} classification task. Consequently, the \gls{dnn} model is trained using the cross-entropy loss:

\begin{equation}
    \label{eqn:xentropy}
    L(P', Y) = \sum_{k=1}^{c+w-1} Y[k] \log(\sigma(P'[k])) + (1 - Y[k]) \log(1 - \sigma(P'[k])),
\end{equation}

where \(\sigma\) denotes the softmax function, applied to the elements of \(P'\) indexed by \(k\) to derive template-match probabilities. \(Y[k]\) represents the elements of the one-hot vector \(Y\), indicating the ground truth window position within the context.

Now, we can train the model exclusively on \(L(P', Y)\), aiming not just to predict window positions accurately but also to jointly-learn the imperfection-invariant, temporally-equivalent representations \(W'\) and \(C'\). Also, this approach supports identifying the top-\(k\) most likely positions post training, laying the groundwork for heuristical score following, discussed in \cref{subsecn:Heuristic.Score.Following}.

Once the prescribed model has been trained, the most probable window position within the context, \(Y_{pred}\), is trivially determined using the operation:

\begin{equation}
    \label{eqn:pred.argmax}
    Y_{pred} = argmax(P')
\end{equation}

Given the inherent unpredictability of performance imperfections, and to ensure that the accompaniment would consistently follow the performer, our approach treats score following and computer accompaniment as iterative processes, executed at predetermined intervals throughout a performance. This fixed rate of execution, \(f_{e}\) Hz, also enables executing the algorithm irrespective of the expected tempo of the solo, enabling us to support following a larger bandwidth of performance tempos.

\subsection{Heuristic Score Following}
\label{subsecn:Heuristic.Score.Following}

To address the challenge of accurately predicting window positions in contexts with similar or repeating note sequences, our approach must extend beyond relying solely on \cref{eqn:pred.argmax}. While some existing solutions consider the entire solo performance up to the current time, this method significantly increases computational demand for longer performances. To mitigate this challenge while keeping \(c\) and \(w\) constant, we combine \cref{eqn:pred.argmax} with a set of heuristic rules based on past predictions. We derive these rules by largely applying common-sense logic that is universally applicable to the flow of performances regardless of the instrument or genre involved.

The heuristic logic for determining the overall predicted score position includes the following steps:

\begin{enumerate}
    \item When the performance consists of repeating patterns, cross-correlation might return similar probabilities for all of them. Therefore, beyond identifying the peak position with \cref{eqn:pred.argmax}, we also consider alternate positions having comparable probabilities. We find these by smoothing the \gls{dnn} output vector, and identifying significant peaks in it using SciPy's \texttt{find\_peaks} function\footnote{\url{https://docs.scipy.org/doc/scipy/reference/generated/scipy.signal.find_peaks.html}} with a prominence threshold of at least 3. If no significant peaks are found, the position of the highest peak of the smoothed vector is selected.

    \item At the start of the performance, the heuristic system might require some time to stabilize and accumulate performance data. During this phase, we only focus on the most prominent peak in the \gls{dnn}'s output, storing initial predictions in a ring buffer. Once sufficient data is accumulated, all significant peaks are evaluated in conjunction with recent data from this buffer.

    \item Using linear regression on buffer data, we extrapolate a predicted position that is consistent with the performer's local tempo, serving as a reference to critique the \gls{dnn} model's predictions. A model prediction is validated if it meets the following criteria:
          \begin{enumerate}
              \item It is greater or slightly less than the previous prediction.
              \item It lies within a defined threshold relative to the buffer prediction.
              \item The rate of change compared to the latest prediction is comparable to the buffer's trend.
          \end{enumerate}
          We derive the first rule by assuming that the performer is almost always going to perform monotonically from the start of the score to its end; the slight threshold of backtracking is to overcome from potential errors in preceding predictions. The second and third rules stem from the expectation that a performer would continue to maintain a rate similar to the most recent performance section (the section being under a second long).

    \item If a valid model prediction is not found, we calculate the mean of the buffer and model predictions;
          \begin{enumerate}
              \item If this mean is close to the model's prediction, it is possible that both the buffer and model predictions might be off. In this case, a safe estimate would be to consider the mean value as the predicted position.
              \item If not, the model prediction has jumped too far, and this estimation may be incorrect. Therefore, we use the buffer predicted value.
          \end{enumerate}

    \item While relying on the buffer predicted position is useful to safeguard against isolated errors made by the model, it could inhibit quick response to changes in the performer's dynamics. Furthermore, errors from the linear interpolation could also accumulate. Thus, we place a limit on the possible number of consecutive buffer predictions. Upon reaching this limit, we assume that the model is responding to a change in the performance characteristics, and use its prediction as the estimated score location\footnote{In this scenario, it would also be useful to perform the global scan approach, discussed in \cref{subsecn:Global.Search.for.Out.of.Context.Performance.Windows}.}.
\end{enumerate}

\begin{figure}[h!]
    \centering
    \includegraphics{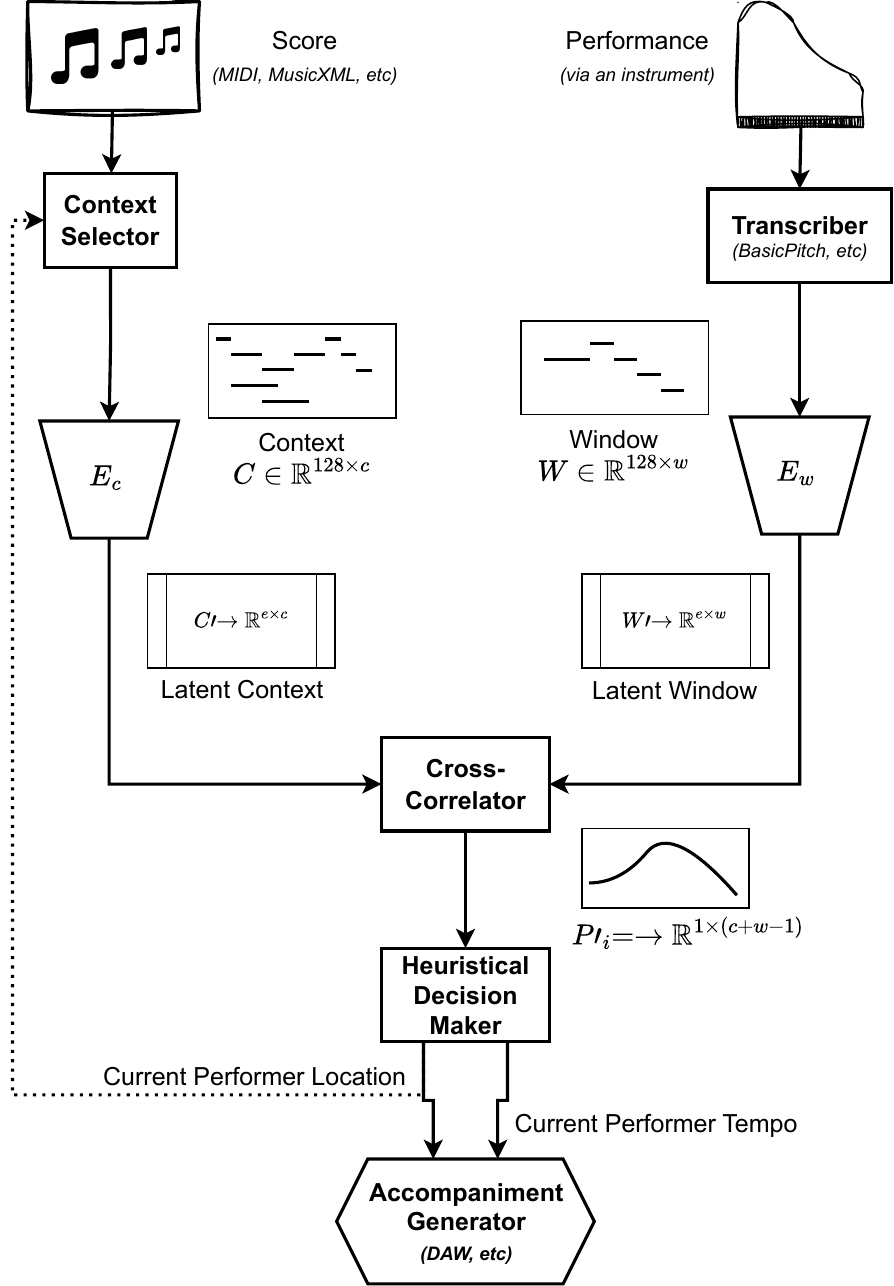}
    \caption{A block diagram of the score following system defined in \cref{secn:problem.defn}.}
    \label{fig:proposed.system.block.diagram}
\end{figure}

These heuristics add a practical layer to our score-following strategy, combining algorithmic estimation with an adaptability characteristic of human-like tracking, thus enhancing the reliability and responsiveness of the accompaniment system to the performer's nuances. The overall score following system is illustrated in \cref{fig:proposed.system.block.diagram}.

\section{Implementation Details}
\label{subsecn:Implementation.Details}

This section delineates the methodologies employed in implementing the problem defined in \cref{secn:problem.defn}. We begin with 
discussing the architecture employed by the score following \gls{dnn}, followed by the dataset and paradigm used for training it. Then, we discuss its integration into a comprehensive system designed for real-time performer tracking and low-latency accompaniment generation.

\subsection{Model Architecture: Tyke}
\label{subsecn:sf.model.archs}

Building upon previously established insights of score progression, we hypothesize that the position of a new window is very likely to be present in the vicinity of the previous window location. This premise suggests that searching within a localized score segment—a smaller context—would be both time and compute efficient. However, significant imperfections in the performance could position the new window entirely outside this context, necessitating searching within a larger section of the score with support from the previously predicted locations. Moreover, given the variable lengths of musical scores, our model must embody temporal-equivariance, as expounded in \cref{subsecn:effective.repr}.

Accordingly, we introduce \textbf{Tyke}, a compact convolutional model designed to pinpoint the window's location within a localized context. Tyke aims to determine the probabilities of the window belonging to each context position (refer to \crefrange{eqn:xcorr}{eqn:pred.argmax} and \cref{subsecn:Heuristic.Score.Following}). Tyke's design also allows for processing windows and contexts of arbritrary durations as long as they exceed its kernel dimensions, facilitating us to use Tyke variants having different values of $c$ without additional training overhead. With this flexibility, we can even search within the entire score by simply setting $c$ to the score length.

\subsection{Training Paradigm}
\label{subsecn:Training.Paradigm}

\subsubsection{Dataset}
\label{subsubsecn:Dataset}

The MAESTRO Dataset V3.0.0 \cite{Hawthorne2018EnablingFP}, comprising approximately 200 hours of virtuosic piano performances with precise alignment ($\sim 3$ ms) between MIDI notes and audio waveforms, forms the basis of our training data. It also includes a CSV file delineating the division of data into training, validation, and testing sets. To train Tyke, we exclusively utilize the MIDI files, processing them to create the \textbf{MAESTRO for Score Following - Static (MSF-S)} dataset as per algorithm~\ref{alg:msfs}. Once generated, the MSF-S samples consists of a tuple of \(C\), \(W\) and \(Y\). During training, we randomly provide these samples as input to Tyke.

\begin{algorithm}
    \caption{Create MSF-S}
    \label{alg:msfs}
    \begin{algorithmic}[1]

        \For{\textbf{each} data split}
        \State Create a CSV to store the split data
        \State Initialize $n \gets 0$
        \While{$n < n_{\text{split}}$}
        \State Select a random MIDI file from the data split
        \State Process the piano instrument to a piano-roll
        \For{\textbf{each} note in piano-roll}
        \If{note velocity $> 0$}
        \State note velocity $\gets 1$
        \Else
        \State note velocity $\gets 0$
        \EndIf
        \EndFor
        \State Determine a random position in the piano roll from where a context of duration $c$ can be obtained
        \State Determine a random window start position
        \If{window is completely outside context}
        \State flag $\gets$ True
        \Else
        \State flag $\gets$ False
        \EndIf
        \State Append to CSV: MIDI file name, context start, window start, flag
        \State $n \gets n + 1$
        \EndWhile
        \EndFor

    \end{algorithmic}
\end{algorithm}

\subsubsection{MIDI Augmentations with MIDIOgre}

To derive imperfection-invariant representations, as detailed in \cref{subsecn:effective.repr}, we developed \textbf{MIDIOgre}\footnote{\url{https://github.com/a-pillay/MIDIOgre}}, a Python-based, MIT-licensed MIDI data augmentation toolkit.  MIDIOgre can parse MIDI files as PrettyMIDI \cite{Raffel2014Intuitive} objects to simulate performance imperfections on the fly, enhancing Tyke's training with a broad spectrum of potential imperfections. In our strategy, these imperfections are introduced randomly, without any grounding on human performers doing the same. This follows common applications of augmentation in the vision domain \cite{Shorten2019ASO}, adhering to the expectation that our model would be asymptotically exposed to the entire set of human-induced imperfections over multiple epochs of training. We hypothesize this not only ensures robustness against unpredictable performance variances but also minimizes overfitting.

We adhere to the common strategies described by \gls{mdtk} \cite{Mcleod2020TheMD} to implement the following augmentations in MIDIOgre:
\begin{enumerate}
    \item \texttt{PitchShift}: Randomly transposes MIDI notes of randomly selected instruments.
    \item \texttt{OnsetTimeShift}: Randomly alters note onset times while maintaining their durations.
    \item \texttt{DurationShift}: Randomly modifies note durations while preserving their onset times.
    \item \texttt{NoteDelete}: Randomly removes some notes from an instrument track.
    \item \texttt{NoteAdd}: Randomly inserts some notes into an instrument track.
\end{enumerate}

For visualising how these augmentations appear within piano rolls, refer \cref{ap:Visualizing.MIDIOgre.Augmentations}.

\subsection{Accompaniment System Design\footnote{Note that this subsection does not introduce new components but suggests ways to seamlessly integrate existing tools and techniques with the score follower discussed in \cref{subsecn:sf.model.archs}.}}
\label{subsubsecn:Auxillary.System.Design}

In this section, we describe potential ways to synergistically combine existing auxillary components to form a comprehensive accompaniment system capable of listening to audio frames of the solo, convert them into piano-rolls, infer them using the score following \gls{dnn}(s), and generate the accompaniment in response to the predicted position of the performer. In this regard, our overarching goal is a system having minimal latency while being sufficiently reactive to be able to track even the most rapidly-played, ephemeral notes.

\subsubsection{Audio to MIDI Conversion via BasicPitch}

Since providing an accompaniment system that should follow analog instruments means we would not readily have the solo in a digital format, the first component in our accompaniment system would involve converting the incoming solo audio frames into a binary piano-roll format that our score following \gls{dl} model (see \cref{subsecn:sf.model.archs}) can process.

To this end, we could employ \textbf{BasicPitch} \cite{Bittner2022ALI}, a low-resource pre-trained \gls{amt} model that generalizes to a wide range of polyphonic instruments (including vocals). BasicPitch has been experimentally validated to have comparable performance to instrument-specific state-of-the-art \gls{amt} systems while being a considerably smaller model. Additionally, BasicPitch is also available as a Python library\footnote{\url{https://github.com/spotify/basic-pitch}} that can be conveniently used to perform \gls{amt} on the provided audio through the \gls{cli}.

As a potential implementation, we first record a frame of the audio solo followed by processing it through suitable noise-filtering and gain conditioning \gls{dsp} blocks. Subequently, we run it through BasicPitch to generate the corresponding piano-roll window that will be supplied to the downstream score following system. For maximizing efficiency, we could also omit BasicPitch's intermediate MIDI file generation process and finetune it to directly generate the window that can be processed as is by Tyke.

\subsubsection{Parallel Processing}
Considering that human auditory perception is sensitive to latencies greater than 10ms \cite{Lelic2022HearingAD}, we estimate a successful computer accompaniment system to have input-to-output latencies under 5ms. From a preliminary analysis, \gls{amt}s like BasicPitch and the score following model can be the main bottlenecks in our system. Additionally, Python's \gls{gil} places heavy restrictions on running all the system components in parallel.

As a workaround, we use the multithreading capabilities of QTThreadPool, available through PySide6\footnote{\url{https://pypi.org/project/PySide6/}}, a Python wrapper for the QT6 graphics toolkit. Specifically, implementing dedicated threads for capturing the audio frames, executing the DNNs, updating the front end GUI and communicating the score position via \gls{osc}. Further, it would be beneficial to implement a shared tensor queue that gets populated by the latest recorded audio frame at a fixed time interval. Then, the \gls{dnn} thread can fetch the latest data of size \(w\) from this queue and provide them as inputs to the model, at a rate of \(f_{e}\) per second.

\subsubsection{GUI design}

To visually validate the progression of score following, we suggest designing a simple graphical front end using PySide 6. The core elements displayed on the front end should include the score and the incoming notes from the solo. The current context, window and the detected window location must also be highlighted. Essentially, the front end serves largely as a qualitative evaluation tool, but can be extended upon to support functionalities as deemed necessary down the line.

\subsubsection{Accompaniment using \gls{osc} and \gls{daw}}
Once the window location within the score is identified, we must use this information to deliver the accompaniment. The \gls{osc} protocol\footnote{\url{https://ccrma.stanford.edu/groups/osc/index.html}} has been widely implemented as a means of exchanging musical information between different performance tools. It is natively supported by most popular DAWs and musical experimentation software like Max/MSP \footnote{\url{https://cycling74.com/products/max}} and SuperCollider \footnote{\url{https://supercollider.github.io/}}. \gls{osc} messages can be communicated as text strings over the \gls{udp} \cite{Postel1980UserDP} protocol. Consequently, our accompaniment component could set up a \gls{udp} client over an available port, and periodically communicate the following information as \gls{osc} messages:
\begin{enumerate}
    \item Estimated position of the solo on the score, as a time stamp.
    \item Estimated tempo of the performer at the given instant (expressed in terms of its deviation from the base tempo set by the score).
\end{enumerate}

We believe that any commonly available \gls{daw} can utilize the information described above to generate an appropriate accompaniment for the solo. A probable impediment in this method is that most DAWs implement their \gls{osc} \gls{api} idiosyncratically. Thus, the system would have to communicate the messages in a format specific to the \gls{daw} being used. However, this added effort only manifests as an additional configuration effort that can be set once when the system is initialized.

Should future research determine that using \gls{daw}s is impractical for computer accompaniment, exploring alternative solutions like Dannenberg's \textit{Accomplice} might be beneficial. Accomplice is an accompaniment system that provides fine control over tempo and score positions and includes functionalities to load, start, and stop projects, along with clock synchronization for precise timing.
\chapter{Evaluation}
\label{ch:Evaluation}

In addition to the development of the system outlined in \cref{ch:Proposed.System}, it is crucial to establish a set of metrics to evaluate its performance reliably, alongside further experimentation and refinement to achieve an optimal level of functionality. As discussed in \cref{subsection:Performance.Metrics}, we aim to consolidate and refine existing evaluation metrics to ensure their relevance and compatibility with both current and future research. This chapter begins with a definition of the metrics selected for our system's assessment, followed by establishing a set of experiments where we benchmark it on these metrics. In addition, we set thresholds of acceptable metric values, aiming to match or improve upon state-of-the-art score followers where comparable metrics are available. In scenarios lacking direct comparisons, we will deduce these thresholds based on general performance expectations from the system.

\section{Training}
\label{secn:Training.Metrics}

In the absence of directly comparable neural frameworks for training Tyke, we utilize the cross-correlation between unencoded windows (\(W\)) and unencoded contexts (\(C\)) as described in \cref{subsecn:Aligning.the.Solo.with.the.Performance}, to establish our \textit{training baseline}\footnote{This should not be confused with our score following baseline, Flippy, introduced in \cref{secn:Inference.Evaluation.Metrics}.}. This approach not only facilitates monitoring of Tyke's training progress and prevent overfitting but also aids in evaluating the effectiveness of the learned latent features in enhancing the score following accuracy of our system.

During the training phase, we monitor the following metrics:
\begin{enumerate}
    \item \textbf{Cross-Entropy Loss} as a proxy to measure the model's classification error.
    \item \textbf{Prediction Accuracy} to quantify the percentage of window locations Tyke correctly estimates within a 5ms range of their actual ground truth values.
    \item \textbf{Baseline Prediction Accuracy} to determine the percentage of window locations accurately estimated by the
    cross-correlation 
    of \(W\) and \(C\)
    within a 5-ms range of their actual ground truth values.
\end{enumerate}

\begin{table}[h!]
    \centering
    \begin{tabular}{|l|cc|}
        \hline
        \multirow{2}{*}{\textbf{Metric}} & \multicolumn{2}{c|}{\textbf{Variable Name}}                                                \\ \cline{2-3}
                                         & \multicolumn{1}{c|}{\textbf{Train Set}}     & \multicolumn{1}{c|}{\textbf{Validation Set}} \\ \hline \hline
        Cross-Entropy Loss               & \multicolumn{1}{c|}{train\_loss}            & \multicolumn{1}{c|}{val\_loss}               \\ \hline
        Prediction Accuracy              & \multicolumn{1}{c|}{train\_acc}             & \multicolumn{1}{c|}{val\_acc}                \\ \hline
        Baseline Prediction Accuracy     & \multicolumn{1}{c|}{-}                      & \multicolumn{1}{c|}{val\_bacc}               \\ \hline
    \end{tabular}%
    \caption{List of training metrics for Tyke.}
    \label{tab:tyketrainingmetrics}
\end{table}

\Cref{tab:tyketrainingmetrics} lists the variables tracking these metrics across the train, validation, and test data splits during Tyke's training process. The following conditions serve as thresholds to ascertain successful training of Tyke:
\begin{enumerate}
    \item To avoid overfitting: The ratio of validation to training accuracy (\(\frac{val\_acc}{train\_acc}\)) should be at least 0.75.
    \item For accurate onset prediction: The validation accuracy (\(val\_acc\)) should meet or exceed 0.9.
    \item To outperform the baseline
    (piano-roll data cross-correlation): Validation accuracy (\(val\_acc\)) should surpass the baseline prediction accuracy (\(val\_bacc\)).
\end{enumerate}

This setup aims to ensure that Tyke not only achieves high accuracy in score following but also surpasses existing methodologies in reliability and efficiency.

\section{Inference Evaluation}
\label{secn:Inference.Evaluation.Metrics}

Following the training of our \gls{dnn} model, we assess its performance in score following by analyzing the model using prerecorded human performance data and corresponding scores. Our primary dataset is the (n)ASAP dataset by Peter et al. \cite{Peter2023AutomaticNS}, chosen for its recency and the richness of its contents. (n)ASAP, derived from the ASAP dataset \cite{Foscarin2020ASAPAD}, consists of note-level annotations and is characterized by significant tempo variations and musical diversity. The original performances, sourced from the MAESTRO dataset \cite{Hawthorne2018EnablingFP}, feature a broad spectrum of performer skill levels \cite{Lee2022MusicalSF}.

For evaluating our work, we select a set of metrics primarily adapted from Lee's 2022 work \cite{Lee2022MusicalSF}, which builds on the \gls{mirex} benchmark \cite{Cont2007EvaluationOR} and the methodologies of Thickstun et al. \cite{Thickstun2020RethinkingEM}. The metrics are defined as follows:
\begin{enumerate}
    \item \textbf{Misalign Rate} (\(r_{e}\)): The percentage of performance events where predicted onset times deviate beyond a misalignment threshold, \(\theta_{e}\), from their expected values. We evaluate performances at thresholds \(\theta_{e} \in \{25, 50, 75, 100, 125, 300, 500, 750, 1000\}\) ms.
    \item \textbf{Alignment Error} (\(e_{i}\)): The absolute time difference between the predicted and actual onsets, reported as mean (\(\mu_{e}\)) \(\pm\) \gls{sd} (\(\sigma_{e}\)) over all aligned events.
    \item \textbf{Latency} (\(l_{i}\)): The delay in recognizing an event by the score follower, expressed as mean (\(\mu_{l}\)) \(\pm\) \gls{sd} (\(\sigma_{l}\)) for the entire performance.
\end{enumerate}

As the \textit{score following baseline}\footnote{The score following baseline differs from the training baseline
described
in \cref{secn:Training.Metrics}. The latter is utilized solely for monitoring Tyke's training progress.}, we utilize the system from Lee's work, henceforth referred to as \textbf{Flippy} following the nomenclature used within its accompanying code repository\footnote{\url{https://github.com/flippy-fyp/flippy}}. Specifically, we employ Flippy in its "NSGT-CQT Online" mode, which features a \gls{nsgt} based feature extractor and an \gls{oltw}-based aligner, operating on frames of performance audio and the synthesized score. Our experiments adhere to the configuration Lee used to report their results, maintaining all system parameters at their default values\footnote{\url{https://github.com/flippy-fyp/flippy/blob/main/lib/args.py}}, except for adjusting the "OLTW Weight to constrain the path for the i direction parameter" to 0.5.

While (n)ASAP contains note-level ground truth alignments between the MusicXML score and MIDI performance data, applying these directly to evaluate both systems is challenging. Thus, we devise the following evaluation methods:
\begin{enumerate}
    \item Our system operates on a window of performance notes to determine its overall location within the score. While it is possible to indirectly determine the positions of the constituent notes within the window, this process is complex and could introduce errors orthogonal to the score following process. This is especially critical when predictions occur at score positions where no new notes are onset, as (n)ASAP contains only note-onset specific alignment data. Consequently, we derive ground truth predictions by performing unconstrained, offline \gls{dtw} between the score and performance piano rolls using the dtaidistance\footnote{\url{https://dtaidistance.readthedocs.io/en/latest/modules/dtw.html\#dtaidistance.dtw.warping\_paths\_fast}} Python package. We then calculate the evaluation metrics as follows:
          \begin{enumerate}
              \item Alignment Error: For window \(i\), performance window position \(X_{i}\), ground truth score position \(Y_{i}\), and predicted score position \(Y_{pred_{i}}\):
                    \begin{equation}
                        Y_{i} = dtw\_warping\_path(X_{i})
                    \end{equation}
                    \begin{equation}
                        e_{i} = |Y_{i} - Y_{pred_{i}}|
                    \end{equation}
              \item Misalign Rate: Derived from \(e_{i}\) and \(\theta_{e}\) as previously defined.
              \item Latency: The difference in wall times measured from the recording of the last performance event to when the system generates a prediction output.
          \end{enumerate}
    \item For Flippy, we adopt the string-matching based ground truth generation tool provided by the authors\footnote{\url{https://github.com/flippy-fyp/flippy-quantitative-testbench}}, for simplicity and consistency with reported results.
\end{enumerate}

Given our \gls{dnn}'s training on the MAESTRO \textit{train} split, it is critical to evaluate our system using (n)ASAP performances from the \textit{test} split of MAESTRO exclusively. Of the 74 eligible performances, we were unable to generate Flippy ground truths for 26 due to the string-matching aligner exceeding Python's recursion limit of 5000. Additionally, 4 more performances were excluded since calculating their DTW warping paths exceeded our computational budget, leaving us with 44 (n)ASAP performances available for evaluation, detailed in \cref{tab:nasap_performances}.

\subsection{Inference Experiments}
\label{subsecn:Inference.Experiments}

Our system relies on cross-correlation for rapid execution at a fixed frequency (\(f_{e}\)). However, this approach is sensitive to the base tempo of the performance. When the performance tempo significantly deviates from the tempo indicated by the score, as commonly observed in \cref{tab:nasap_performances}, the resulting mismatch in the resolutions of the corresponding piano rolls can compromise the efficacy of the scale-sensitive cross-correlation algorithm. To understand the impact of these mismatches, we initially evaluate our system under ideal conditions by adjusting the score's base tempo to closely match the estimated tempo of the performance. We then analyze performance with the original score tempos to measure the effect of tempo mismatches. Additionally, we focus on a specific performance (P7) to explore how varying degrees of tempo mismatch influence our system's score following capability. This analysis involves plotting the misalignment rate (\(r_{e}\)) across a range of tempo mismatches.

Furthermore, we examine the robustness of our heuristic system by analyzing the influence of hyperparameter values on its performance. Specifically, we vary \(f_{e}\) while keeping all other hyperparameters constant, assessing the system’s behavior on performance P8 at a fixed \(\theta_{e} = 100\) ms.

\begin{table}[]
    \centering
    \resizebox{\columnwidth}{!}{%
        \begin{tabular}{|l|l|c|c|c|}
            \hline
            \begin{tabular}[c]{@{}l@{}}Performance\\ Code\end{tabular} & Performance                                  & \begin{tabular}[c]{@{}c@{}}Duration\\ (s)\end{tabular} & \begin{tabular}[c]{@{}c@{}}Original Score Tempo\\ (BPM)\end{tabular} & \begin{tabular}[c]{@{}c@{}}Performance Tempo\\ (BPM)\end{tabular} \\ \hline\hline
            P1                                                         & Bach/Fugue/bwv\_858/VuV01M                   & 144.75                                                 & 120                                                                  & 58.02                                                             \\ \hline
            P2                                                         & Bach/Prelude/bwv\_858/VuV01M                 & 80.9                                                   & 120                                                                  & 66.74                                                             \\ \hline
            P3                                                         & Bach/Fugue/bwv\_858/Zhang01M                 & 129.19                                                 & 120                                                                  & 65.01                                                             \\ \hline
            P4                                                         & Bach/Prelude/bwv\_858/Zhang01M               & 87.21                                                  & 120                                                                  & 61.91                                                             \\ \hline
            P5                                                         & Bach/Fugue/bwv\_862/Song04M                  & 155.01                                                 & 120                                                                  & 54.18                                                             \\ \hline
            P6                                                         & Bach/Prelude/bwv\_862/Song04M                & 75.19                                                  & 120                                                                  & 105.32                                                            \\ \hline
            P7                                                         & Bach/Fugue/bwv\_863/LeeN01M                  & 169.17                                                 & 120                                                                  & 58.16                                                             \\ \hline
            P8                                                         & Bach/Prelude/bwv\_863/LeeN01M                & 102.88                                                 & 120                                                                  & 50.73                                                             \\ \hline
            P9                                                         & Bach/Fugue/bwv\_863/Shychko01M               & 194.03                                                 & 120                                                                  & 50.71                                                             \\ \hline
            P10                                                        & Bach/Prelude/bwv\_863/Shychko01M             & 102.03                                                 & 120                                                                  & 51.15                                                             \\ \hline
            P11                                                        & Bach/Fugue/bwv\_863/TongB01M                 & 173.55                                                 & 120                                                                  & 56.69                                                             \\ \hline
            P12                                                        & Bach/Prelude/bwv\_863/TongB01M               & 94.35                                                  & 120                                                                  & 55.31                                                             \\ \hline
            P13                                                        & Bach/Fugue/bwv\_873/Lisiecki13M              & 122.01                                                 & 120                                                                  & 103.63                                                            \\ \hline
            P14                                                        & Bach/Prelude/bwv\_873/Lisiecki13M            & 257.22                                                 & 120                                                                  & 64.38                                                             \\ \hline
            P15                                                        & Bach/Fugue/bwv\_891/Duepree06M               & 218.25                                                 & 120                                                                  & 166.59                                                            \\ \hline
            P16                                                        & Bach/Prelude/bwv\_891/Duepree06M             & 141.73                                                 & 120                                                                  & 140.54                                                            \\ \hline
            P17                                                        & Bach/Prelude/bwv\_891/BLINOV04M              & 155.61                                                 & 120                                                                  & 128                                                               \\ \hline
            P18                                                        & Beethoven/Piano\_Sonatas/10-1/Hou02M         & 278.94                                                 & 98                                                                   & 85.87                                                             \\ \hline
            P19                                                        & Beethoven/Piano\_Sonatas/12-1/Kleisen05M     & 406.6                                                  & 56                                                                   & 48.55                                                             \\ \hline
            P20                                                        & Beethoven/Piano\_Sonatas/16-1/BuiJL02M       & 281.44                                                 & 132                                                                  & 138.04                                                            \\ \hline
            P21                                                        & Beethoven/Piano\_Sonatas/16-1/Khmara05M      & 283.75                                                 & 132                                                                  & 136.91                                                            \\ \hline
            P22                                                        & Beethoven/Piano\_Sonatas/16-1/LeeSH02M       & 306.07                                                 & 132                                                                  & 126.93                                                            \\ \hline
            P23                                                        & Beethoven/Piano\_Sonatas/16-1/LuoJ03M        & 264.95                                                 & 132                                                                  & 146.63                                                            \\ \hline
            P24                                                        & Beethoven/Piano\_Sonatas/16-1/Woo05M         & 277.57                                                 & 132                                                                  & 139.96                                                            \\ \hline
            P25                                                        & Beethoven/Piano\_Sonatas/18-1/ChenGuang03M   & 365.41                                                 & 138                                                                  & 108.79                                                            \\ \hline
            P26                                                        & Beethoven/Piano\_Sonatas/18-1/LeungR02M      & 354.07                                                 & 138                                                                  & 112.27                                                            \\ \hline
            P27                                                        & Beethoven/Piano\_Sonatas/18-1/Levitsky05M    & 352.44                                                 & 138                                                                  & 112.79                                                            \\ \hline
            P28                                                        & Beethoven/Piano\_Sonatas/18-1/ZhangH05M      & 352.48                                                 & 138                                                                  & 112.78                                                            \\ \hline
            P29                                                        & Beethoven/Piano\_Sonatas/5-1/Colafelice02M   & 272.96                                                 & 197                                                                  & 186.18                                                            \\ \hline
            P30                                                        & Beethoven/Piano\_Sonatas/5-1/SunD02M         & 237.09                                                 & 197                                                                  & 214.34                                                            \\ \hline
            P31                                                        & Beethoven/Piano\_Sonatas/8-3/Na06M           & 258.88                                                 & 190                                                                  & 194.22                                                            \\ \hline
            P32                                                        & Beethoven/Piano\_Sonatas/9-1/Tysman05M       & 434.06                                                 & 148                                                                  & 89.29                                                             \\ \hline
            P33                                                        & Beethoven/Piano\_Sonatas/9-3/Tysman05M       & 194.31                                                 & 168                                                                  & 161.49                                                            \\ \hline
            P34                                                        & Chopin/Etudes\_op\_10/12/Bult-ItoS04M        & 156.18                                                 & 160                                                                  & 126.88                                                            \\ \hline
            P35                                                        & Chopin/Etudes\_op\_10/12/HuNY03M             & 158.07                                                 & 160                                                                  & 125.36                                                            \\ \hline
            P36                                                        & Chopin/Etudes\_op\_10/12/LuoJ08M             & 136.2                                                  & 160                                                                  & 145.5                                                             \\ \hline
            P37                                                        & Chopin/Etudes\_op\_10/12/WuuE06M             & 150.65                                                 & 160                                                                  & 131.54                                                            \\ \hline
            P38                                                        & Chopin/Etudes\_op\_10/12/ZhangYunling02M     & 138.18                                                 & 160                                                                  & 143.41                                                            \\ \hline
            P39                                                        & Chopin/Sonata\_2/4th/KaszoS16M               & 81.88                                                  & 220                                                                  & 225.69                                                            \\ \hline
            P40                                                        & Debussy/Pour\_le\_Piano/1/MunA12M            & 225.49                                                 & 152                                                                  & 143.05                                                            \\ \hline
            P41                                                        & Liszt/Concert\_Etude\_S145/1/Kleisen06M      & 220.05                                                 & 120                                                                  & 105.79                                                            \\ \hline
            P42                                                        & Liszt/Concert\_Etude\_S145/1/Woo06M          & 236.24                                                 & 120                                                                  & 98.54                                                             \\ \hline
            P43                                                        & Liszt/Concert\_Etude\_S145/2/Lu03M           & 155.91                                                 & 240                                                                  & 213.97                                                            \\ \hline
            P44                                                        & Rachmaninoff/Preludes\_op\_23/6/Nikiforov14M & 170.9                                                  & 72                                                                   & 60.38                                                             \\ \hline
        \end{tabular}%
    }
    \caption{Shortlisted performances from the (n)ASAP dataset used for evaluating our system and the Flippy baseline. Observations from the two rightmost columns indicate a significant mismatch between the score-specified tempos and the actual performance tempos. As part of our evaluation experiments, we study the impact of these tempo mismatches on our system’s score following capabilities.}
    \label{tab:nasap_performances}
\end{table}

\section{Ablation Study}
\label{secn:Ablations}

To thoroughly benchmark the MIDI data augmentation strategies described in \cref{subsecn:Training.Paradigm}, we also conduct an ablation study. This study will compare the performance of our model with and without each specific augmentation technique to ascertain their individual contributions to the model's effectiveness. The study involves training multiple variations of the best performing Tyke architecture, each progressively disabling one of the augmentations detailed in \cref{subsecn:Training.Paradigm}. We will also include a model variant trained entirely without augmentations
for comparison. This series of ablations is designed to study the impact of each augmentation strategy on developing robust latent features that are essential for effective score following.

\newpage

\section{Listening Evaluation}
\label{secn:Listening.Evaluation.Guidelines}

Beyond numerical metrics, we also perform listening tests on the performances listed in \cref{tab:nasap_performances} to evaluate the efficacy of our score follower from a holistic perspective. This allows us to assess the system's suitability for following performers in practice, identifying both strengths and areas needing improvement. We perform these tests by warping the score piano roll onto the performance using paths returned by the \gls{dtw} ground truth and our score follower. We then synthesize these piano rolls using a \gls{daw}, and play the performance and the score through separate channels. Listening via separate ears helps us visualize how the evaluation metrics manifest in real life, and how well the system handles different kinds of performance deviations.

\chapter{Results}
\label{ch:results}

\section{Training}
\label{secn:Res.Training.Results}

With respect to the metrics defined in \cref{secn:Training.Metrics}, the \textbf{MiniTyke} architecture, as described in \cref{tab:tyke.architecture}, achieved optimal performance. This model incorporates a single 1D-\gls{cnn} layer for \(E_{c}\) and \(E_{w}\), followed by a \gls{relu} activation layer. By setting \(e\) to 64, we compress the resultant latent representations \(C'\) and \(W'\) to half their original size, thereby simplifying downstream cross-correlation computations. MiniTyke is a remarkably compact \gls{dnn}, with fewer than 50,000 parameters, thus incurring minimal computational costs during inference. We set the piano roll resolution to \(\frac{1}{96}\)s, with \(c = 512\) and \(w = 256\), corresponding to context durations of approximately 5.33s and window durations of approximately 2.67s, respectively. Training employed the AdamW optimizer with a weight decay of \(1e-2\), a learning rate of \(5e-4\), and the default values for \(\beta_{1}\) and \(\beta_{2}\). Additionally, a Cosine Annealing learning rate scheduler was utilized with a minimum learning rate of \(1e-6\) and a quarter-cycle of 10 epochs. The training spanned 50 epochs, with each epoch comprising 500 training samples and 50 validation samples from the MSFS-S dataset described in \cref{subsecn:Training.Paradigm}. To model performer imperfections, we applied MIDIOgre augmentations on \(W\) using the configurations outlined in \cref{tab:minityke_midiogre}, identified following a qualitative analysis of human performances relative to their scores.

The batch size was set to 64, and the model was trained on a single NVIDIA GeForce RTX 3060 Mobile GPU over a duration of 67 minutes. Optimal results were achieved by the 45th epoch, as detailed in \cref{tab:minityke_training_performance}. These results fulfill all objectives established in \cref{secn:Training.Metrics}, confirming the successful training of the MiniTyke model and its readiness for evaluation within the overall score following framework. Furthermore, the \(6\%\) improvement in validation accuracy over the
training
baseline cross-correlation algorithm indicates that the latent features learned by MiniTyke not only facilitate data compression but might also enhance the system’s score following accuracy.

\begin{table}[h!]
    \centering
    \begin{tabular}{|l|c|c|c|}
        \hline
        Layer (type)                                       & Output Shape                & Param \# & Details                             \\
        \hline\hline
        \textbf{MiniTyke}                                  & [\(B\), 767]                & --       & --                                  \\
        \hline
        \textit{Sequential \(E_{c}\):}                     & [\(B\), 64, 512]            & --       & --                                  \\
        \indent Conv1d-1                                   & [\(B\), 64, 512]            & 24,640   & kernel\_size=3, stride=1, padding=1 \\
        \indent ReLU-2                                     & [\(B\), 64, 512]            & --       & --                                  \\
        \hline
        \textit{Sequential \(E_{w}\):}                     & [\(B\), 64, 256]            & --       & --                                  \\
        \indent Conv1d-3                                   & [\(B\), 64, 256]            & 24,640   & kernel\_size=3, stride=1, padding=1 \\
        \indent ReLU-4                                     & [\(B\), 64, 256]            & --       & --                                  \\
        \hline
        \multicolumn{2}{|c|}{\textbf{Total params}}        & \multicolumn{2}{c|}{49,280}                                                  \\
        \hline
        \multicolumn{2}{|c|}{\textbf{Total mult-adds (M)}} & \multicolumn{2}{c|}{605.55}                                                  \\
        \hline
    \end{tabular}
    \caption{Detailed summary of the \gls{dnn} architecture for MiniTyke for \(c = 512\) and \(w = 256\).}
    \label{tab:tyke.architecture}
\end{table}

\begin{table}[h!]
    \centering
    \resizebox{\columnwidth}{!}{%
        \begin{tabular}{|l|c|c|c|}
            \hline
            \textbf{Augmentation} & \textbf{Configuration}                                                                                                                                                 & \textbf{mode} & \textbf{Probability} \\ \hline\hline
            PitchShift            & \texttt{max\_shift = 5}                                                                                                                                                & both          & \multirow{5}{*}{0.1} \\ \cline{1-3}
            OnsetTimeShift        & \texttt{max\_shift = 0.5}                                                                                                                                              & both          &                      \\ \cline{1-3}
            DurationShift         & \texttt{max\_shift = 0.25}                                                                                                                                             & both          &                      \\ \cline{1-3}
            NoteDelete            & -                                                                                                                                                                      & -             &                      \\ \cline{1-3}
            NoteDelete            & \begin{tabular}[c]{@{}c@{}}\texttt{note\_num\_range}=(20, 120)\\ \texttt{note\_duration\_range}=(0.5, 1.5)\\ \texttt{restrict\_to\_instrument\_time}=True\end{tabular} & -             &                      \\ \hline
        \end{tabular}%
    }
    \caption{MIDIOgre augmentations used to train MiniTyke, with corresponding configurations used.}
    \label{tab:minityke_midiogre}
\end{table}

\begin{table}[h!]
    \centering
    \resizebox{\columnwidth}{!}{%
        \begin{tabular}{|l|c|c|c|c|c|}
            \hline
            \textbf{Model} & \textbf{train\_loss} & \textbf{val\_loss} & \textbf{train\_acc} & \textbf{val\_acc} & \textbf{val\_bacc} \\ \hline\hline
            MiniTyke       & 1.341                & 1.458              & 86\%                & 94\%              & 88\%               \\ \hline
        \end{tabular}%
    }
    \caption{Best training performance for MiniTyke.}
    \label{tab:minityke_training_performance}
\end{table}

\section{Inference Evaluation}
\label{secn:Res.Inference.Evaluation}

After training the MiniTyke model, we integrate it with the heuristics described in \cref{subsecn:Heuristic.Score.Following} to form a complete score following system, termed \textbf{HeurMiT}. We optimized HeurMiT using recalibrated scores, achieving the best results with settings \(f_{e} = 10\)Hz, $w = 500$ (\(\sim 5.21\)s), and $c = 1250$ (\(\sim 13.02\)s). To smooth MiniTyke's output, we applied a moving average filter with a window size of 5 and set the ring buffer size to store the past 20 predictions, with the first 5 marking the stabilization phase. For a MiniTyke prediction to be considered valid, it must fall within -48 to +96 samples of the buffer predicted position, and the rate of change with respect to the last prediction must be between 0.5 and 1.5. The maximum number of consecutive buffer predictions allowed is set to 5. For evaluating Flippy, we adopted the default settings prescribed by Lee.

Utilizing the metrics outlined in \cref{secn:Inference.Evaluation.Metrics}, we quantified HeurMiT's performance against the baseline system developed by Lee on valid instances from the (n)ASAP dataset. Our findings are presented in \cref{tab:HeurMiT_vs_flippy_sf_eval_metrics}. Flippy exhibited a better misalign rate (\(r_{e}\)) across all values of the misalignment threshold (\(\theta_{e}\)). For alignment errors (\(e_{i}\)), HeurMiT performed slightly better than Flippy at \(\theta_{e} < 125\) ms, but this trend reversed at higher values. These results indicate that HeurMiT has more tendency to lose track of the performance, as evidenced by the higher values of \(r_{e}\). In real-life, we want minimal \(e_{i}\) throughout the performance and in these situations, HeurMiT follows the performer more accurately than Flippy, provided it does not completely lose track. High values of \(e_{i}\) at larger \(\theta_{e}\) could stem from HeurMiT not having safeguards against the window drifting completely outside the context due to accumulated errors in past predictions. In these cases, the cross-correlation might return an incorrect position with the highest likelihood\footnote{We discuss potential solutions to this issue in \cref{subsecn:Global.Search.for.Out.of.Context.Performance.Windows}.}, with heuristic rules attempting to establish a reasonable compromise.

A notable difference between the systems lies in their latency (\(l\)), with HeurMiT apparently outperforming Flippy by an order of \(10^3\). However, we refrain from claiming substantial gains in efficiency due to this disparity arising from the fundamentally different operational methodologies of the two systems. Flippy conducts \gls{oltw} between the spectral representations of the entire synthesized score and the available performance audio up to that point, which constitutes an audio-to-audio alignment algorithm with a complexity of \(\mathcal{O}(\max(p, s))\), where \(p\) and \(s\) represent the lengths of the performance and the score, respectively. In contrast, HeurMiT utilizes lightweight cross-correlation on fixed-length piano-roll windows and contexts, yielding an \(\mathcal{O}(1)\) algorithm. This is further enhanced by using Pytorch's \texttt{nn.functional.conv1d}\footnote{\url{https://pytorch.org/docs/stable/generated/torch.nn.functional.conv1d.html}}, which includes computational optimizations for NVIDIA GPUs.

For a visual comparison of HeurMiT's predictions with the \gls{dtw} ground truth for select performances, see \cref{ap:Tyke.Inference.Evaluation.Plots}.

\begin{table}[h!]
    \centering
    \resizebox{\columnwidth}{!}{%
        \begin{tabular}{lcccccc}
            \hline
            \multicolumn{1}{|l|}{\multirow{2}{*}{\begin{tabular}[c]{@{}l@{}}Misalignment\\ Threshold [\(\theta_{e}\)]\\ (ms)\end{tabular}}} & \multicolumn{3}{c|}{Flippy (Lee \cite{Lee2022MusicalSF})}                                                               & \multicolumn{3}{c|}{HeurMiT (ours)}                                                                                                                                                                                                                                                                                                                                                                                                                                                                                                                                                                                 \\ \cline{2-7}
            \multicolumn{1}{|l|}{}                                                                                                                      & \multicolumn{1}{c|}{\begin{tabular}[c]{@{}c@{}}Misalign\\ Rate [\(r_{e}\)]\\ (\%)\end{tabular}} & \multicolumn{1}{c|}{\begin{tabular}[c]{@{}c@{}}Alignment\\ Error [\(e_{i}\)]\\ (ms)\end{tabular}} & \multicolumn{1}{c|}{\begin{tabular}[c]{@{}c@{}}Latency [\(l\)]\\ (ms)\end{tabular}} & \multicolumn{1}{c|}{\begin{tabular}[c]{@{}c@{}}Misalign\\ Rate [\(r_{e}\)]\\ (\%)\end{tabular}} & \multicolumn{1}{c|}{\begin{tabular}[c]{@{}c@{}}Alignment\\ Error [\(e_{i}\)]\\ (ms)\end{tabular}} & \multicolumn{1}{c|}{\begin{tabular}[c]{@{}c@{}}Latency [\(l\)]\\ (ms)\end{tabular}}                                                      \\ \hline\hline
            \multicolumn{1}{|l|}{25}                                                                                                                    & \multicolumn{1}{c|}{\textbf{83.12}}                                                                                     & \multicolumn{1}{c|}{11.31 \(\pm\) 12.3}                                                                                    & \multicolumn{1}{c|}{1505.89 \(\pm\) 509.99}                                         & \multicolumn{1}{c|}{92.99}                                                                                               & \multicolumn{1}{c|}{\textbf{10.58 \(\pm\) 7.28}}                                                                           & \multicolumn{1}{c|}{\multirow{9}{*}{\textbf{\begin{tabular}[c]{@{}c@{}}1.1 \(\pm\) 0.19 (cuda)\\ 6.07 \(\pm\) 1.70 (cpu)\end{tabular}}}} \\ \cline{1-6}
            \multicolumn{1}{|l|}{50}                                                                                                                    & \multicolumn{1}{c|}{\textbf{70.63}}                                                                                     & \multicolumn{1}{c|}{22.47 \(\pm\) 23.88}                                                                                   & \multicolumn{1}{c|}{1547.46 \(\pm\) 544.11}                                         & \multicolumn{1}{c|}{88.07}                                                                                               & \multicolumn{1}{c|}{\textbf{19.94 \(\pm\) 13.08}}                                                                          & \multicolumn{1}{c|}{}                                                                                                                    \\ \cline{1-6}
            \multicolumn{1}{|l|}{75}                                                                                                                    & \multicolumn{1}{c|}{\textbf{60.82}}                                                                                     & \multicolumn{1}{c|}{33.38 \(\pm\) 33.27}                                                                                   & \multicolumn{1}{c|}{1586.03 \(\pm\) 596.42}                                         & \multicolumn{1}{c|}{80.97}                                                                                               & \multicolumn{1}{c|}{\textbf{33.37 \(\pm\) 21.52}}                                                                          & \multicolumn{1}{c|}{}                                                                                                                    \\ \cline{1-6}
            \multicolumn{1}{|l|}{100}                                                                                                                   & \multicolumn{1}{c|}{\textbf{54.05}}                                                                                     & \multicolumn{1}{c|}{43.69 \(\pm\) 42.23}                                                                                   & \multicolumn{1}{c|}{1678.77 \(\pm\) 729.09}                                         & \multicolumn{1}{c|}{77.24}                                                                                               & \multicolumn{1}{c|}{\textbf{40.91 \(\pm\) 26.74}}                                                                          & \multicolumn{1}{c|}{}                                                                                                                    \\ \cline{1-6}
            \multicolumn{1}{|l|}{125}                                                                                                                   & \multicolumn{1}{c|}{\textbf{49.66}}                                                                                     & \multicolumn{1}{c|}{51.9 \(\pm\) 53.54}                                                                                    & \multicolumn{1}{c|}{1636.95 \(\pm\) 781.11}                                         & \multicolumn{1}{c|}{74.11}                                                                                               & \multicolumn{1}{c|}{\textbf{50.88 \(\pm\) 31.34}}                                                                          & \multicolumn{1}{c|}{}                                                                                                                    \\ \cline{1-6}
            \multicolumn{1}{|l|}{300}                                                                                                                   & \multicolumn{1}{c|}{\textbf{40.93}}                                                                                     & \multicolumn{1}{c|}{\textbf{83.15 \(\pm\) 90.84}}                                                                          & \multicolumn{1}{c|}{1459.31 \(\pm\) 879.93}                                         & \multicolumn{1}{c|}{59.97}                                                                                               & \multicolumn{1}{c|}{107.59 \(\pm\) 77.35}                                                                                  & \multicolumn{1}{c|}{}                                                                                                                    \\ \cline{1-6}
            \multicolumn{1}{|l|}{500}                                                                                                                   & \multicolumn{1}{c|}{\textbf{37.06}}                                                                                     & \multicolumn{1}{c|}{\textbf{122.83 \(\pm\) 130.8}}                                                                         & \multicolumn{1}{c|}{1467.38 \(\pm\) 942.84}                                         & \multicolumn{1}{c|}{54.35}                                                                                               & \multicolumn{1}{c|}{159.67 \(\pm\) 121.48}                                                                                 & \multicolumn{1}{c|}{}                                                                                                                    \\ \cline{1-6}
            \multicolumn{1}{|l|}{750}                                                                                                                   & \multicolumn{1}{c|}{\textbf{34.39}}                                                                                     & \multicolumn{1}{c|}{\textbf{154.2 \(\pm\) 171.99}}                                                                         & \multicolumn{1}{c|}{1445.23 \(\pm\) 957.16}                                         & \multicolumn{1}{c|}{50.38}                                                                                               & \multicolumn{1}{c|}{220.54 \(\pm\) 177.89}                                                                                 & \multicolumn{1}{c|}{}                                                                                                                    \\ \cline{1-6}
            \multicolumn{1}{|l|}{1000}                                                                                                                  & \multicolumn{1}{c|}{\textbf{32.48}}                                                                                     & \multicolumn{1}{c|}{\textbf{186 \(\pm\) 215.02}}                                                                           & \multicolumn{1}{c|}{1441.31 \(\pm\) 1005.59}                                        & \multicolumn{1}{c|}{47.18}                                                                                               & \multicolumn{1}{c|}{284.34 \(\pm\) 240.03}                                                                                 & \multicolumn{1}{c|}{}                                                                                                                    \\ \hline
                                                                                                                                                        & \multicolumn{1}{l}{}                                                                                                    & \multicolumn{1}{l}{}                                                                                                       & \multicolumn{1}{l}{}                                                                & \multicolumn{1}{l}{}                                                                                                     & \multicolumn{1}{l}{}                                                                                                       & \multicolumn{1}{l}{}
        \end{tabular}%
    }
    \caption{Comparison of score following inference evaluation metrics for HeurMiT vs. Flippy. For all metrics, lower values indicate better performance.}
    \label{tab:HeurMiT_vs_flippy_sf_eval_metrics}
\end{table}

\begin{figure}[h!]
    \centering
    \includegraphics{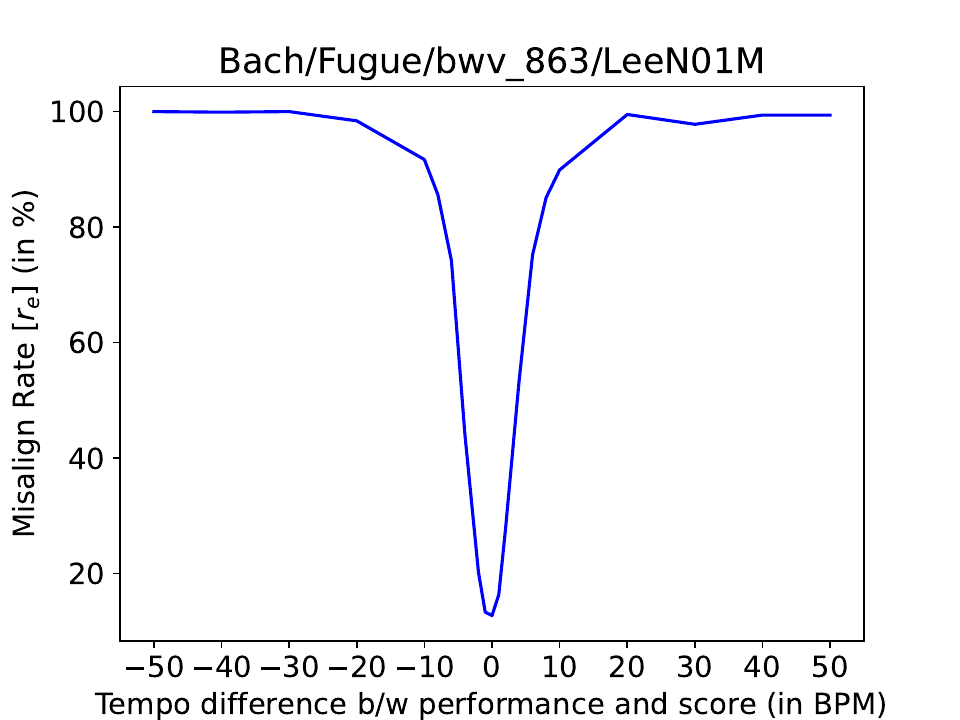}
    \caption{Impact of varying performance-score tempo mismatches on HeurMiT's misalign rate (\(r_{e}\)) for P7 at \(\theta_{e} = 100\)ms.}
    \label{fig:tc_vs_re}
\end{figure}

\Cref{tab:heurmitoriginaltempos} presents the results when HeurMiT attempts to follow performances with tempo mismatches relative to the score. The extremely high values of \(r_{e}\) across all \(\theta_{e}\) thresholds indicate that HeurMiT struggles to adequately follow these performances. Further analysis is illustrated in \cref{fig:tc_vs_re}, where we plot the impact of varying degrees of performance-score tempo mismatches for performance P7. We observe that HeurMiT's \(r_{e}\) exceeds beyond \(50\%\) as the mismatch exceeds \(\pm5\) BPM. This poor performance can be attributed to the non scale-invariant nature of cross-correlation and the small kernel size used in MiniTyke, which only considers 3 consecutive samples at a time. This critical limitation compromises the robustness of HeurMiT, which was intended to adapt to significant changes in performance tempo.

\begin{table}[h!]
    \centering
    \begin{tabular}{|l|c|c|}
        \hline
        \begin{tabular}[c]{@{}l@{}}Misalignment\\ Threshold [\(\theta_{e}\)]\\ (ms)\end{tabular} & \begin{tabular}[c]{@{}c@{}}Misalign\\ Rate [\(r_{e}\)]\\ (\%)\end{tabular} & \begin{tabular}[c]{@{}c@{}}Alignment\\ Error [\(e_{i}\)]\\ (ms)\end{tabular} \\ \hline\hline
        25                                                                                                   & 99                                                                                                  & 8.95 \(\pm\) 5.95                                                                                     \\ \hline
        50                                                                                                   & 98                                                                                                  & 19.92 \(\pm\) 10.08                                                                                   \\ \hline
        75                                                                                                   & 96                                                                                                  & 36.16 \(\pm\) 20.43                                                                                   \\ \hline
        100                                                                                                  & 95                                                                                                  & 45.19 \(\pm\) 24.49                                                                                   \\ \hline
        125                                                                                                  & 94                                                                                                  & 66.77 \(\pm\) 28.41                                                                                   \\ \hline
        300                                                                                                  & 89                                                                                                  & 145.27 \(\pm\) 77.28                                                                                  \\ \hline
        500                                                                                                  & 85                                                                                                  & 227.00 \(\pm\) 131.01                                                                                 \\ \hline
        750                                                                                                  & 82                                                                                                  & 331.53 \(\pm\) 194.55                                                                                 \\ \hline
        1000                                                                                                 & 78                                                                                                  & 435.14 \(\pm\) 268.79                                                                                 \\ \hline
    \end{tabular}%
    \caption{Inference evaluation metrics for HeurMiT when there is a mismatch between the performance score tempos.}
    \label{tab:heurmitoriginaltempos}
\end{table}

\begin{figure}[h!]
    \centering
    \includegraphics{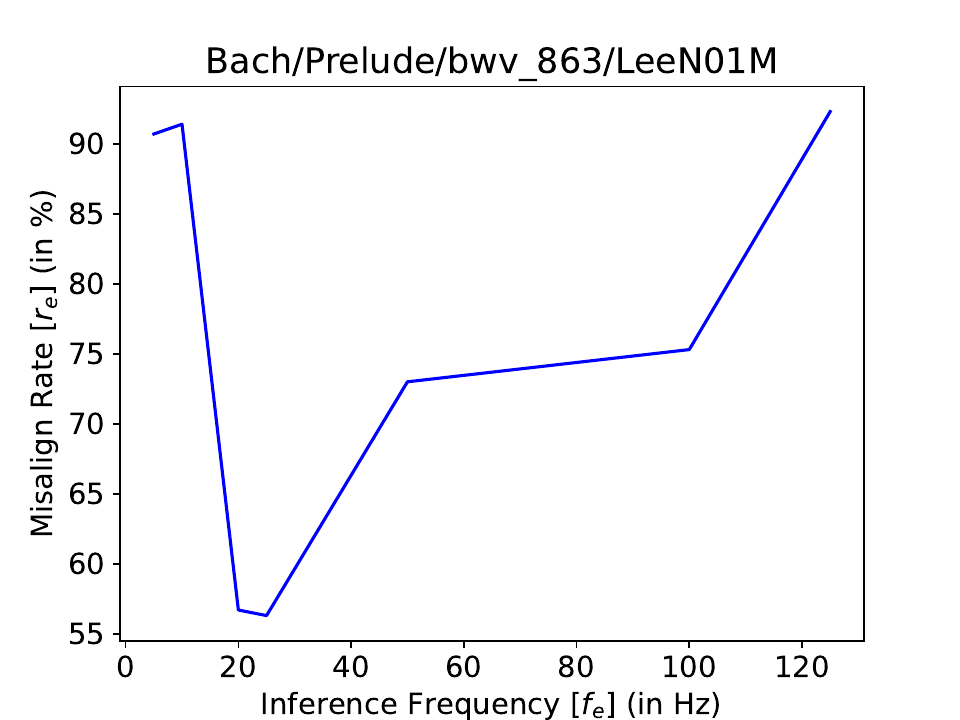}
    \caption{Impact of varying inference frequency (\(f_e\)) on HeurMiT's misalign rate (\(r_{e}\)) for P8 at \(\theta_{e} = 100\)ms.}
    \label{fig:fe_vs_re}
\end{figure}

\Cref{fig:fe_vs_re} shows how HeurMiT's misalign rate (\(r_{e}\)) varies with the inference frequency (\(f_{e}\)), ranging from \(91.4\%\) to \(56.3\%\) at \(\theta_{e} = 100\)ms. This variability indicates that HeurMiT's ability to follow performances is significantly influenced by the applied hyperparameters. Ideally, a robust score follower should not exhibit strong dependency on hyperparameter settings, as this necessitates adjustments for different pieces or performers, which is not favorable.

\section{Ablation Study}
\label{secn:Res.Ablations}

For our ablation study on MIDIOgre augmentations, we trained MiniTyke models using the same configurations as defined in \cref{secn:Res.Training.Results}, but over 25 epochs. The results of this study are presented in \cref{tab:heurmitablations}. Across all ablation experiments, the values of \(r_{e}\) and \(e_{i}\) obtained are quite similar, and the comparative differences can be considered negligible. Consequently, the MIDI augmentations applied do not appear to offer any significant benefits to HeurMiT's robustness.

\begin{table}[h!]
    \centering
    \begin{tabular}{|l|c|c|}
        \hline
        Ablation                                                              & \begin{tabular}[c]{@{}c@{}}Misalign\\ Rate [\(r_{e}\)]\\ (\%)\end{tabular} & \begin{tabular}[c]{@{}c@{}}Alignment\\ Error [\(e_{i}\)]\\ (ms)\end{tabular} \\ \hline\hline
        All 5 Augmentations                                                   & 76                                                                                                  & 43.92 \(\pm\) 26.90                                                                                   \\ \hline
        disabling \texttt{NoteAdd}                                            & 74                                                                                                  & 45.70 \(\pm\) 28.07                                                                                   \\ \hline
        disabling \texttt{NoteDelete}                                         & 76                                                                                                  & 45.04 \(\pm\) 27.57                                                                                   \\ \hline
        disabling \texttt{DurationShift}                                      & 76                                                                                                  & 43.58 \(\pm\) 27.48                                                                                   \\ \hline
        disabling \texttt{OnsetTimeShift}                                     & 76                                                                                                  & 43.87 \(\pm\) 26.93                                                                                   \\ \hline
        disabling \texttt{PitchShift} (\emph{i.e.}, no augmentations applied) & 74                                                                                                  & 44.33 \(\pm\) 26.85                                                                                   \\ \hline
    \end{tabular}%
    \caption{Inference evaluation metrics for HeurMiT upon ablating the applied MIDIOgre augmentations during training. In the first experiment, we apply all five augmentations as described in \cref{tab:minityke_midiogre}. Subsequent experiments progressively disable these augmentations; the second experiment omits \texttt{NoteAdd}, and this pattern continues until the final ablation, where all MIDIOgre augmentations are disabled.}
    \label{tab:heurmitablations}
\end{table}

\section{Listening Evaluation}
\label{secn:Res.Listening.Evaluation}

In our listening evaluation, detailed observations were made on a subset of performances listed in \cref{tab:nasap_performances}, providing insights into the practical quality of score following by HeurMiT. Notably, performances P7, P11, P21, P38, and P43 were followed effectively from start to finish. However, a perceivable lag was noted, particularly during fast note progressions, as in P11, P21, and P43. This lag, consistent across performances, could potentially be corrected by offsetting HeurMiT's predictions by the average lag observed.

Conversely, in cases like P25, P39, and P41, HeurMiT lost track almost immediately, likely because the heuristic system was still in its stabilization phase, lacking sufficient historical data to accurately assess the predictions from the cross-correlation algorithm. A potential solution could involve setting a limit on the maximum allowable increment in score predictions; however, further empirical testing is needed to confirm its effectiveness.

Despite the heuristic protections, there were instances where HeurMiT incorrectly estimated positions in pieces with repetitive note patterns.For P18, HeurMiT was thrown off seeing such patterns at two spots in the performance, but was able to recover quickly since they were followed by a complex note sequence in both cases. In P20, HeurMiT confused sections with similar progressions composed of slightly different notes. Similarly for P34, it jumped to another position in the vicinity of the ground truth having the exact same pattern.

\section{Summary}
\label{secn:Res.Summary}

Under ideal circumstances, evaluation results might suggest that HeurMiT performs comparably to the recently developed Flippy system by Lee. However, these findings are tempered by several limitations:
\begin{enumerate}
    \item HeurMiT performs optimally when the performance tempo closely matches the score, or at a predetermined alternate rate. This is problematic in practice as performers often vary their tempo significantly. The lack of scale invariance in the cross-correlation method used by HeurMiT means that substantial tempo mismatches lead to poor performance.
    \item Variations in hyperparameter settings significantly affect HeurMiT's performance, indicating a reliance on specific performance nuances. This reliance contradicts the goal of developing general-purpose heuristics and suggests a continuation of performance-dependent behaviors observed in other systems.
    \item Unfortunately, our ablation studies do not reveal any distinct benefits from applying the MIDIOgre augmentations towards enhancing HeurMiT's robustness. This outcome, however, could be attributed to the generally subpar performance of our current system. We believe it would be beneficial to reperform our ablations with a better-performing system to more accurately assess the true contributions of the MIDI augmentations described in this work.
    \item Even when inference metrics might suggest adequate performance, listening evaluations reveal that deviations as small as \(\mu_{e} \approx 43\)ms can still be perceptible and may influence performer behavior, potentially leading to a feedback loop of error accumulation.
\end{enumerate}

These observations highlight the need for continued research into robust, efficient score following techniques that can operate independently of specific performance characteristics. Future work must explore alternative approaches to address the shortcomings identified within HeurMiT.

\chapter{Future Directions}
\label{ch:Future.Directions}

Following the identified shortcomings of HeurMiT as discussed in \cref{secn:Res.Summary}, this chapter outlines potential methods for addressing these issues. Initially, we will present straightforward enhancements to the current cross-correlation and heuristics-based approach. Subsequently, we explore alternative \gls{dl}-based methodologies that may entirely circumvent the limitations of the template matching paradigm.

\section{Improvements to the Current Approach}
\label{secn:Improvements.to.the.Current.Approach}

Cross-correlation is advantageous due to its fixed cost \(\mathcal{O}(1)\) nature and its integration within a straightforward \gls{dl} paradigm. We propose several enhancements to mitigate its inherent limitations and improve the robustness of the accompanying heuristics.

\subsection{Cross-Correlation only on Note Onsets}
\label{subsecn:Cross.Correlation.only.on.Note.Onsets}

As identified in \cref{secn:Res.Listening.Evaluation}, mismatches in note durations between performances and scores are common. Although HeurMiT attempts to address this with MiniTyke trained on varied note durations using MIDIOgre, an alternative approach is to entirely disregard note durations and focus solely on their onsets. This method aligns with techniques used in symbol-matching systems where notes are represented as symbols, independent of their duration (\cref{secn:early.work}). By excluding note duration information, we aim to enhance the robustness of cross-correlation, especially when latent representations fail to adjust for duration mismatches. Additionally, this approach could slightly increase the tolerance for tempo variations, as it eliminates resolution discrepancies related to note offsets in the piano rolls.
To implement this, we could truncate the note durations in the piano-roll to a single column width.

\subsection{Multi-Scale Cross-Correlation}
\label{subsecn:Multi.scale.Cross.Correlation}

The scale-invariant limitations of cross-correlation were highlighted in \cref{secn:Res.Summary}, particularly when facing significant tempo changes. Inspired by the Viola-Jones method in face detection \cite{Viola2001RapidOD}, we consider employing a set of parallel multi-scale cross-correlation operations. Each operation would apply a different resolution of the score piano roll on the performance piano roll. Leveraging MiniTyke's capability for batch tensor operations, this approach is anticipated to remain computationally feasible.

Upon determining the peak values across these operations, their magnitudes can serve as a confidence metric to select the most appropriate resolution and its corresponding position prediction. This method also permits integration of historical resolution data into heuristic confidence estimates, which could incorporate significant improvements but requires extensive empirical validation.

\subsection{Global Search for Out-of-Context Performance Windows}
\label{subsecn:Global.Search.for.Out.of.Context.Performance.Windows}

A robust score follower should accommodate unexpected performance deviations that might lead to out-of-context windows, which current cross-correlation methods fail to indicate. This inability can cause prediction errors to accumulate rapidly, resulting in the system losing track of the performance. To mitigate this, we propose a \textit{Global Search} method that recalibrates the context by searching the entire score length when the current window appears out-of-context. Although this method incurs an \(\mathcal{O}(s)\) cost, it is intended to run infrequently, thus minimally adding to the overall computational cost.

Triggering this global search could be managed through heuristic rules after a certain number of consecutive buffer predictions or by incorporating a binary classifier within the Tyke framework. This classifier would operate on the cross-correlation vector \(P'\) to determine if the window is out-of-context, potentially using a simple fully-connected layer trained with Binary Cross-Entropy loss.

\subsection{Incorporating Dynamic Heuristics}
\label{subsecn:Incorporating.Dynamic.Heuristics}

Instead of relying on static, performance-independent heuristics, it may be advantageous to explore methodologies that can learn and adapt to the dynamics of a performance. For example, our current system estimates the performer's tempo through linear regression on past predictions, which may fall short in situations where the performer frequently varies their tempo. In this direction, Xia et al. developed models that learn the interplay of musical expression between two collaborating performers \cite{Xia2015SpectralLF}, demonstrating that these systems outperform simple linear regression for mutual tempo estimation. Ideally, extending this adaptive approach to capture commonly observed performance dynamics between human soloists and accompanists could lead to the complete replacement of the rule-based heuristics described in \cref{subsecn:Heuristic.Score.Following}.

\section{Exploring Alternate \gls{dl}-Based Paradigms}
\label{secn:Exploring.Alternate.DL.Based.Paradigms}

Our current system exhibits sensitivity to tempo mismatches between the performance and score, and also relies heavily on heuristic rules. This section discusses potential \gls{dl} paradigms that could overcome these limitations.

Research in \cref{ch:Literature.Review} highlights methods that follow notes as they come, and can inherently adjust to varying tempos. Exploring \gls{dnn}s with \gls{lstm} layers, trained on sequence-alignment prediction paradigms, offers a promising approach in this direction. Such networks could learn score embeddings and compare them against incoming performance notes to predict positions directly, potentially bypassing the need for heuristic adjustments. Incorporating attention mechanisms, either within \gls{lstm}s or through Transformers \cite{Vaswani2017AttentionIA}, could further enhance efficiency by focusing only on relevant historical data.

Implementing these methods requires a dataset that includes performances, scores, and precise note-level ground truth alignments. The (n)ASAP dataset, published subsequent to our initial approach, could be adapted for this purpose. However, the computational intensity of \gls{lstm}s and Transformers might necessitate compromises between real-time performance capabilities and system robustness during both training and inference.

\section{Training Improvements}
\label{secn:Training.Improvements}

Enhancing the training data quality and diversity can also significantly improve the system's performance.

\subsection{Further Exploration of MIDI Augmentations}
\label{subsecn:Further.Exploration.of.MIDI.Augmentations}

Observations from \cref{secn:Res.Listening.Evaluation} indicate random deviations between performances and scores, which MIDIOgre can synthetically replicate during training. Future research could explore:

\begin{enumerate}
    \item \textbf{Intensifying MIDIOgre Augmentations}: Current training uses a random transformation probability of 0.1 for MIDIOgre, based on qualitative analyses of MAESTRO performances (\cref{tab:minityke_midiogre}). Quantitative research into the optimal configurations for MIDIOgre augmentations could refine our training approach.
    \item \textbf{Extending MIDIOgre with New Augmentations}: Performers sometimes introduce note splits and trills not indicated in the score, leading to inconsistencies in HeurMiT's performance. Implementing \texttt{RandomNoteSplit} and \texttt{RandomTrill} functions could help the system better handle these variations. The former would randomly divide some notes into shorter segments, while the latter would need careful consideration of the typical trill characteristics around long notes.
\end{enumerate}

\subsection{Training On a Wider Variety of Performances}
\label{subsecn:Training.On.a.Wider.Variety.of.Performances}

Although HeurMiT is designed to follow a broad range of musical styles and instruments, MiniTyke training has been confined to classical piano performances from the MAESTRO dataset. To truly enhance HeurMiT’s versatility, it is crucial to expand the training datasets to include a variety of instruments and vocal performances. This would involve collecting new datasets that mirror the comprehensive nature of MAESTRO but encompass a wider range of musical expressions and styles. Ideally, these datasets would provide both audio and MIDI recordings along with accurate alignments, facilitating robust training and evaluation of score following systems across diverse musical categories.

\chapter{Conclusions}
\label{ch:Conclusion}

In this work, we introduced the challenges of score following and computer accompaniment as areas ripe for innovation. We reviewed existing methodologies, highlighting both their salient features and inherent limitations.

Following this review, we introduced a novel framework for score following, HeurMiT, which utilizes the capabilities of \gls{dnn}s. This system is designed to interpret compressed latent feature representations from the score and performance, employing cross-correlation techniques enhanced with practical heuristics to accurately locate the performer's position within the score. To make these representations robust against the performer's deviations from the score, we developed a MIDI data augmentation toolkit and a template-matching based training paradigm, complete with a dataset derived from MAESTRO. Additionally, we outlined a comprehensive set of metrics and experiments for benchmarking our system against existing solutions and potential future developments.

Under ideal scenarios, HeurMiT delivers a score-following performance that is comparable to the existing Flippy baseline system developed by Lee \cite{Lee2022MusicalSF}, but is an \(\mathcal{O}(1)\) system that is orders of magnitude more efficient in terms of computational time. However, subsequent experiments also revealed glaring limitations. Our system struggles with significant and unknown tempo deviations between the performance and the score. The underlying set of heuristics is sensitive to performance-specific nuances and requires manual adjustment for each performance to achieve optimal results. Further, listening evaluations on the performance audio and a warped version of the score derived from HeurMiT's predictions showed that the system tends to have a perceivable lag in following performances, often needing an offset adjustment. Moreover, HeurMiT tends to completely lose track of some performances early on during its stabilization phase, or gets confused among similar looking or repeating sequences of notes.

Following the analysis of these pitfalls, we discussed future research directions that either seek to enhance the template-matching paradigm or explore alternate approaches using \gls{dl} that are independent of the tempo sensitivity issues observed with HeurMiT.

In conclusion, while our work in its present form is not yet ready for real-world application in score following, it represents a significant exploration into the potential of \gls{dl}-based neural score following systems. By presenting our methodology and detailing both the strengths and limitations of our work, we aim to inspire future research towards identifying more robust and efficient methods for tracking performances across a diverse array of instruments, genres, and styles using \gls{dnn}s. We view the challenge of score following and computer accompaniment not merely as a technical hurdle, but as an opportunity to fundamentally enhance musical collaboration, learning, and creativity through the power of computers. By effectively addressing these challenges, we aim to empower musicians and learners alike, broadening the scope of musical expression and collaboration.


\appendix

\chapter{Visualizing MIDIOgre Augmentations}
\label{ap:Visualizing.MIDIOgre.Augmentations}

This chapter provides a visual representation of the effects of MIDIOgre augmentations within the piano roll domain. To illustrate these augmentations, we selected an excerpt from Beethoven's Bagatelle No. 25, commonly known as \textit{Für Elise}. The following plots will demonstrate how various MIDIOgre transformations alter the original musical piece, providing insights into how these augmentations might influence the training of our deep learning models by simulating diverse performance variations.

\begin{minipage}{\textwidth}
   \centering
   \includegraphics[width=.8\linewidth]{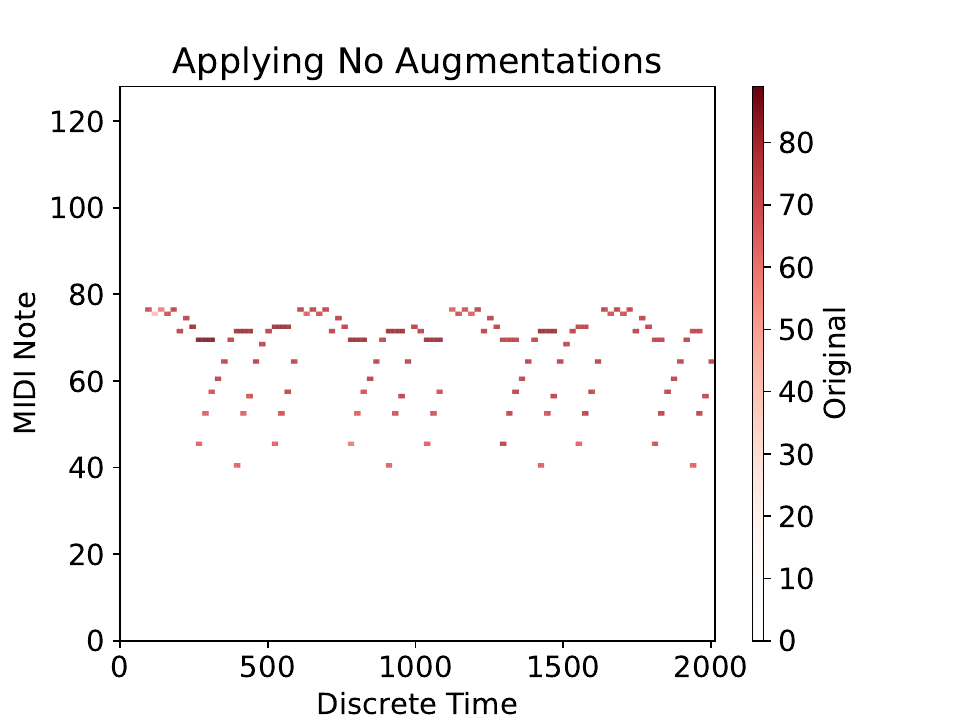}
   \captionof{figure}{Original performance without any MIDIOgre augmentations applied.}
   \label{fig:plot1}
\end{minipage} \\

\begin{table}[ht]
   \centering
   \begin{tabular}{c}

      \begin{minipage}{\textwidth}
         \centering
         \includegraphics[width=.8\linewidth]{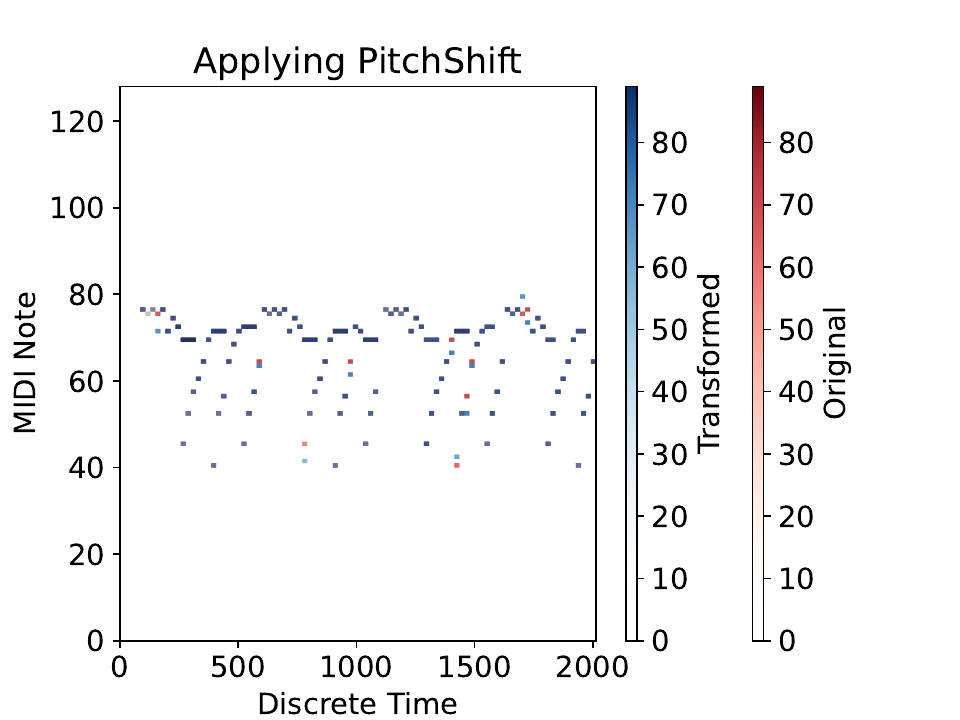}
         \captionof{figure}{Original performance vs PitchShift transforation; \texttt{PitchShift(max\_shift=5, mode='both', p=0.1)}}
         \label{fig:plot1}
      \end{minipage} \\

      \begin{minipage}{\textwidth}
         \centering
         \includegraphics[width=.8\linewidth]{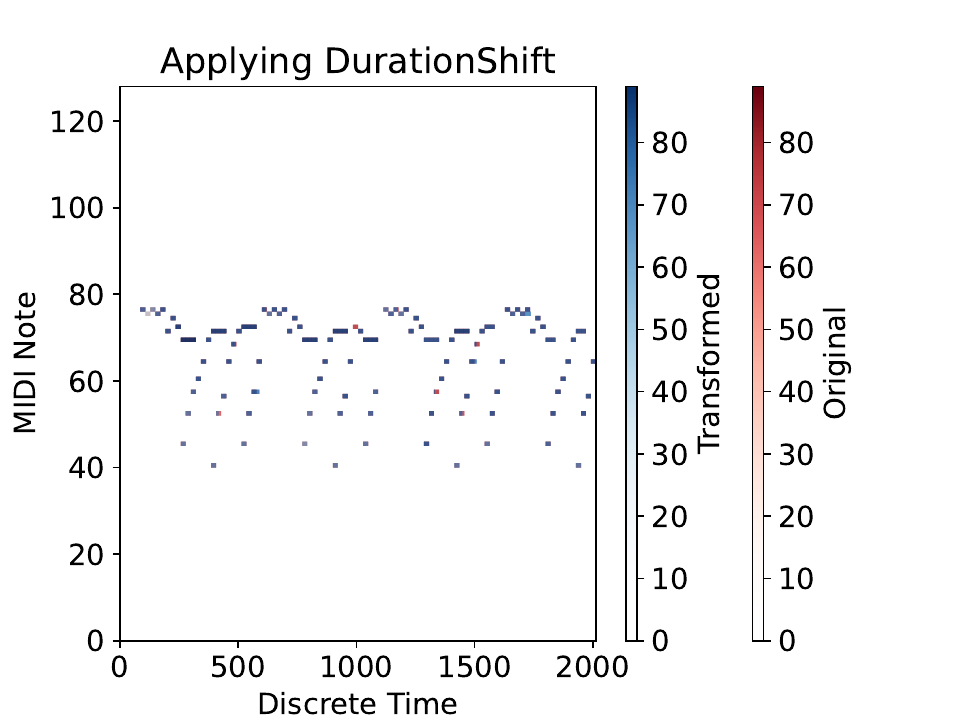}
         \captionof{figure}{\texttt{DurationShift(max\_shift=0.25, mode='both', p=0.1)}}
         \label{fig:plot2}
      \end{minipage}                                        \\
   \end{tabular}
\end{table}

\begin{table}[ht]
   \centering
   \begin{tabular}{c}

      \begin{minipage}{\textwidth}
         \centering
         \includegraphics[width=.8\linewidth]{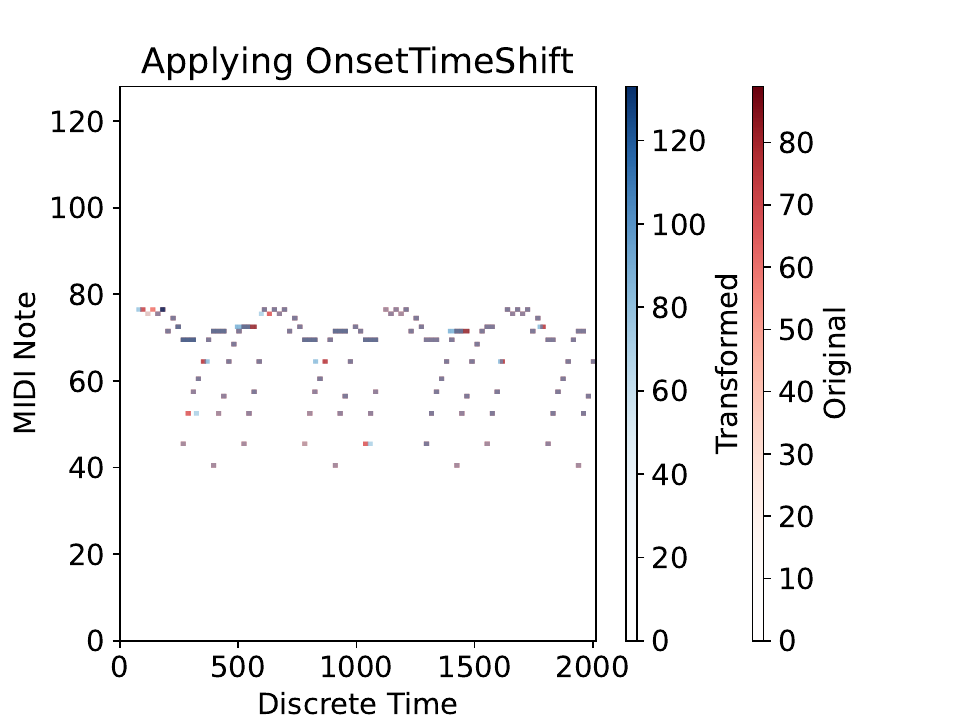}
         \captionof{figure}{\texttt{OnsetTimeShift(max\_shift=0.5, mode='both', p=0.1)}}
         \label{fig:plot2}
      \end{minipage} \\

      \begin{minipage}{\textwidth}
         \centering
         \includegraphics[width=.8\linewidth]{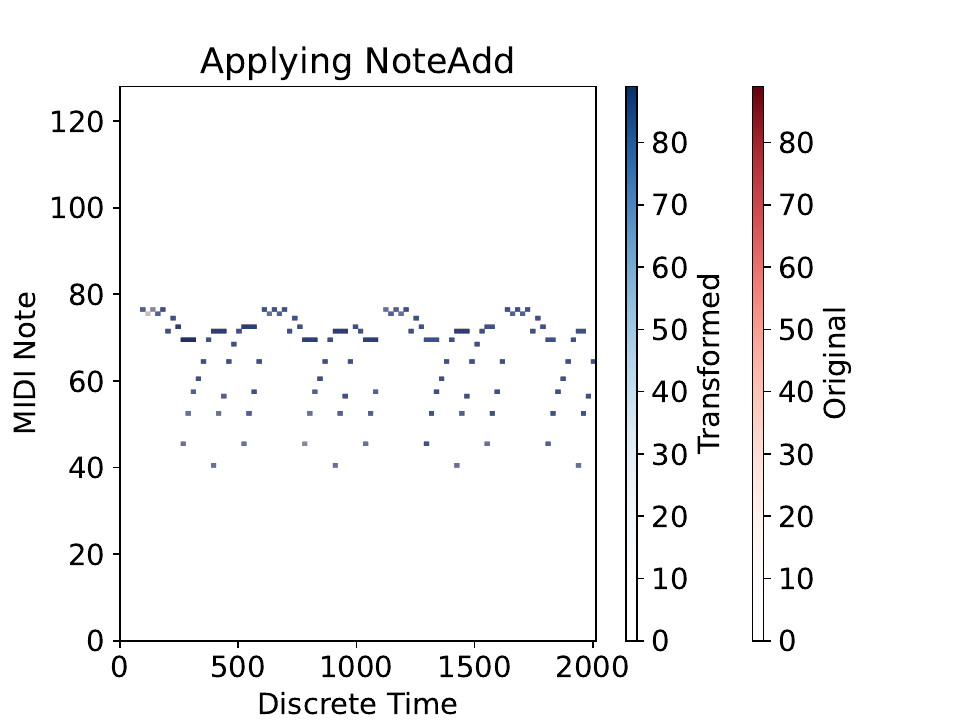}
         \captionof{figure}{\texttt{NoteAdd(note\_num\_range=(25, 120), note\_duration\_range=(0.5, 1.5), restrict\_to\_instrument\_time=True, p=0.1),}}
         \label{fig:plot2}
      \end{minipage}  \\
   \end{tabular}
\end{table}

\begin{table}[ht]
   \centering
   \begin{tabular}{c}

      \begin{minipage}{\textwidth}
         \centering
         \includegraphics[width=.8\linewidth]{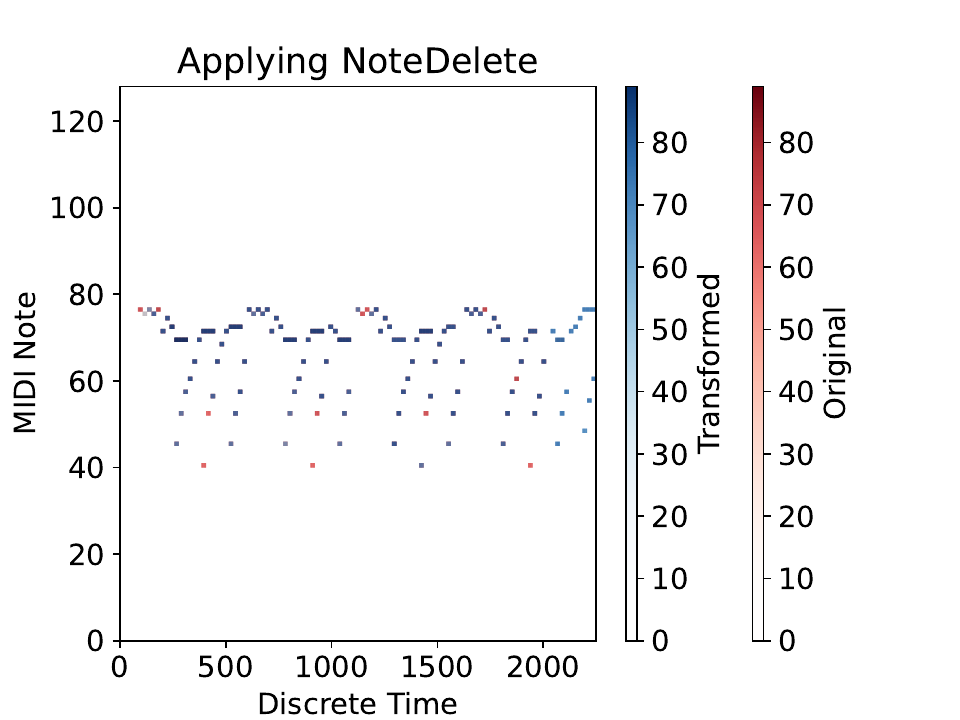}
         \captionof{figure}{\texttt{NoteDelete(p=0.1)}}
         \label{fig:plot2}
      \end{minipage}      \\

      \begin{minipage}{\textwidth}
         \centering
         \includegraphics[width=.8\linewidth]{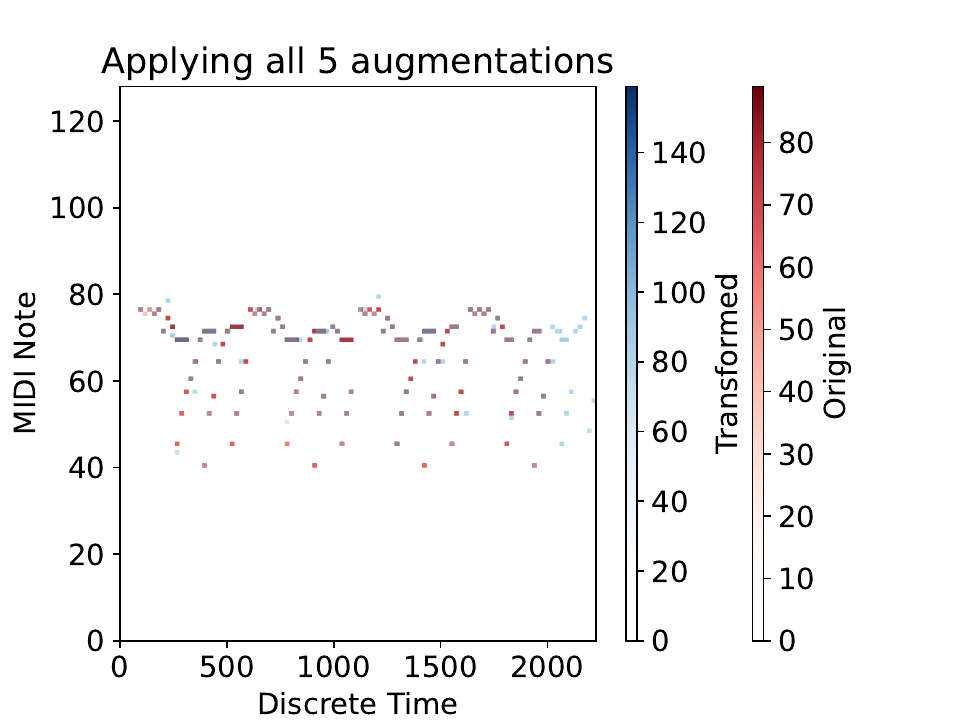}
         \captionof{figure}{All 5 augmentations applied together.}
         \label{fig:plot2}
      \end{minipage} \\
   \end{tabular}
\end{table}


\chapter{Tyke Inference Evaluation Plots}
\label{ap:Tyke.Inference.Evaluation.Plots}

This chapter presents visualizations that illustrate the performance of HeurMiT's score following capabilities, displayed in blue, in comparison to the ground truth DTW warping paths, shown in green, for selected (n)ASAP performances. For a detailed explanation of the methodologies used to compute these metrics, please refer to \cref{secn:Inference.Evaluation.Metrics}. Further analysis of these observations is available in \cref{secn:Res.Inference.Evaluation}. Each plot is labeled with the corresponding performance code as listed in \cref{tab:nasap_performances}.

\begin{table}[ht]
   \centering
   \begin{tabular}{cc}
      \begin{minipage}{.5\textwidth}
         \centering
         \includegraphics[width=.9\linewidth]{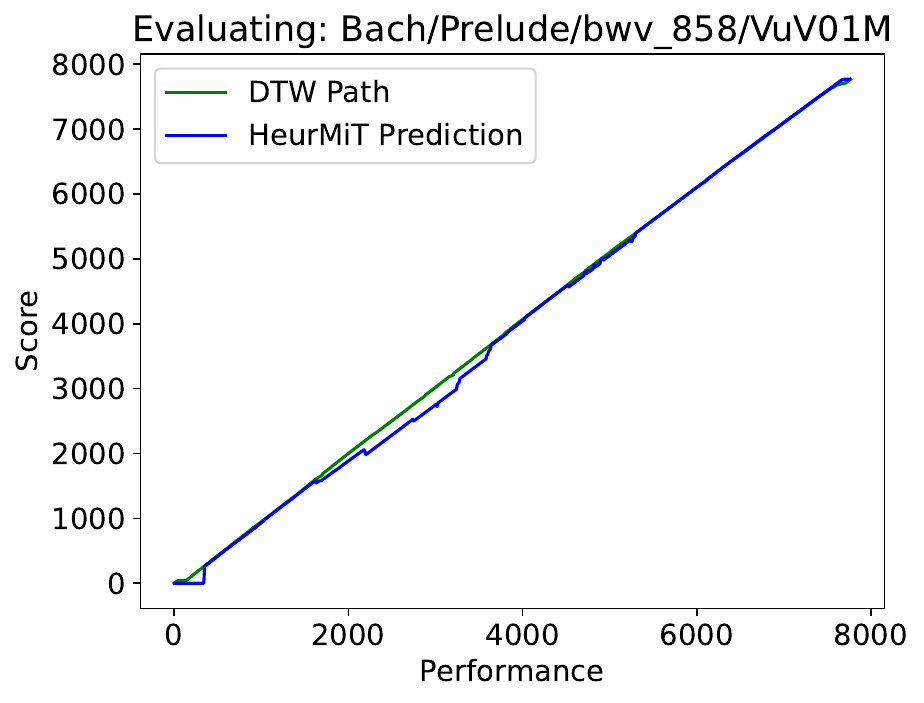}
         \captionof{figure}{P2}
         \label{fig:figure1}
      \end{minipage} &
      \begin{minipage}{.5\textwidth}
         \centering
         \includegraphics[width=.9\linewidth]{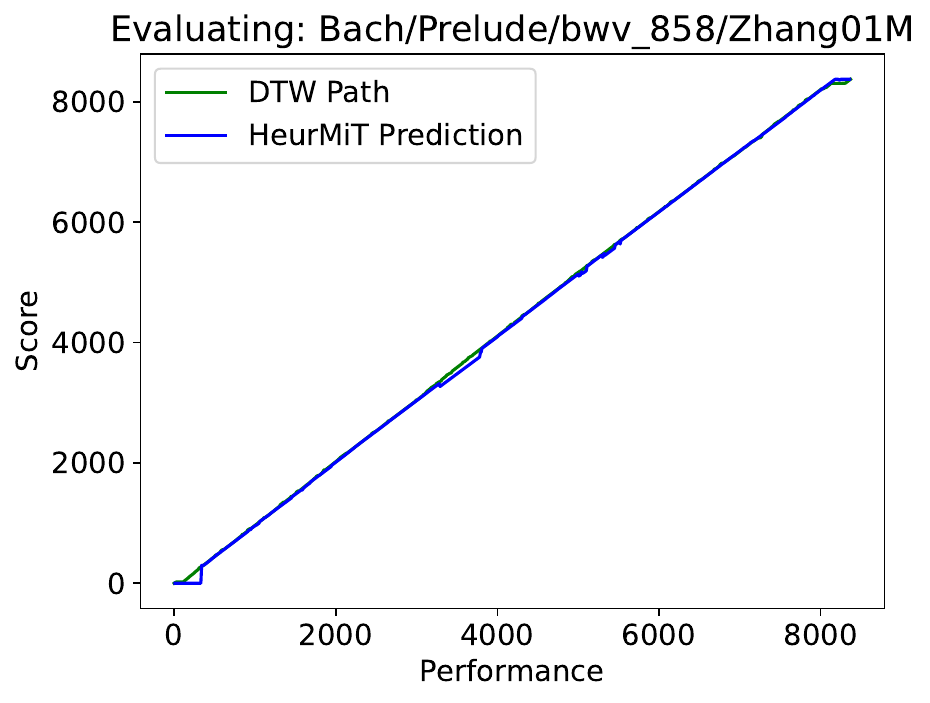}
         \captionof{figure}{P4}
         \label{fig:figure2}
      \end{minipage} \\
   \end{tabular}
\end{table}

\begin{table}[ht]
   \centering
   \begin{tabular}{cc}

      \begin{minipage}{.5\textwidth}
         \centering
         \includegraphics[width=.9\linewidth]{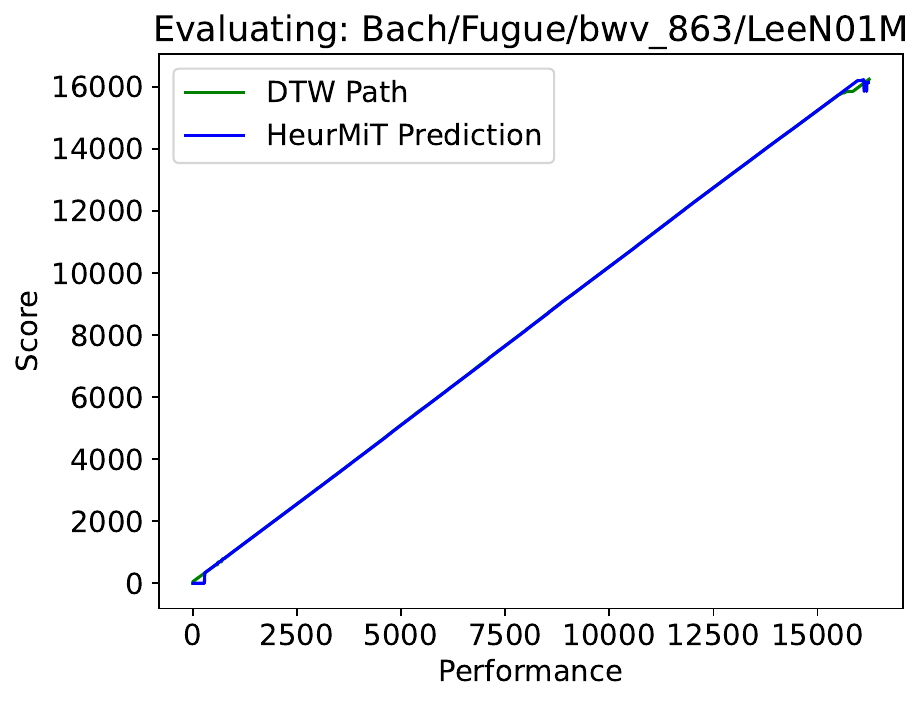}
         \captionof{figure}{P7}
         \label{fig:figure3}
      \end{minipage}  &
      \begin{minipage}{.5\textwidth}
         \centering
         \includegraphics[width=.9\linewidth]{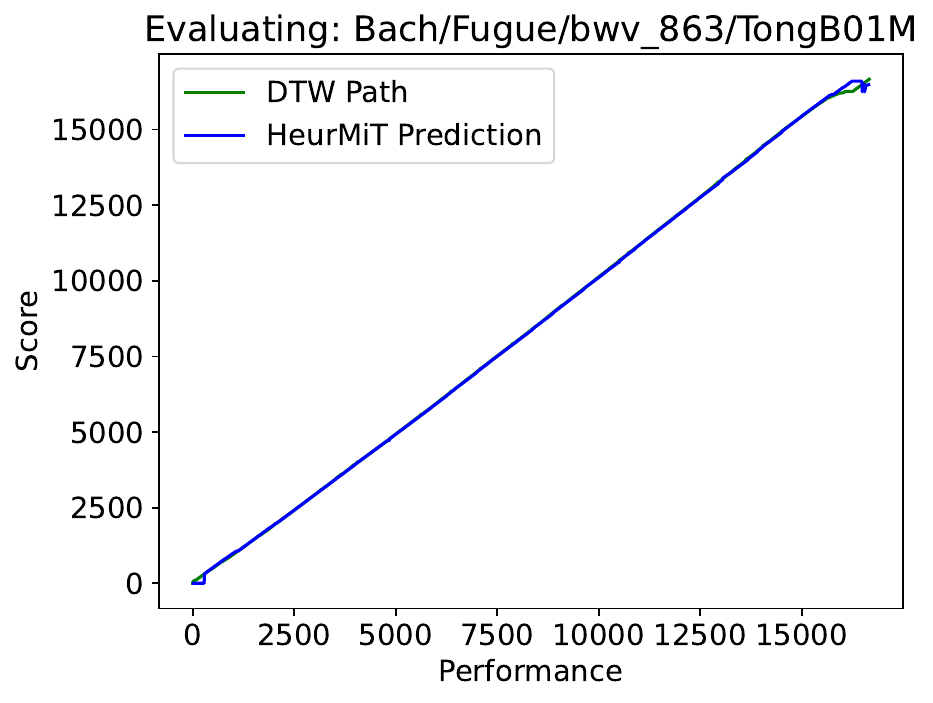}
         \captionof{figure}{P11}
         \label{fig:figure4}
      \end{minipage} \\

      \begin{minipage}{.5\textwidth}
         \centering
         \includegraphics[width=.9\linewidth]{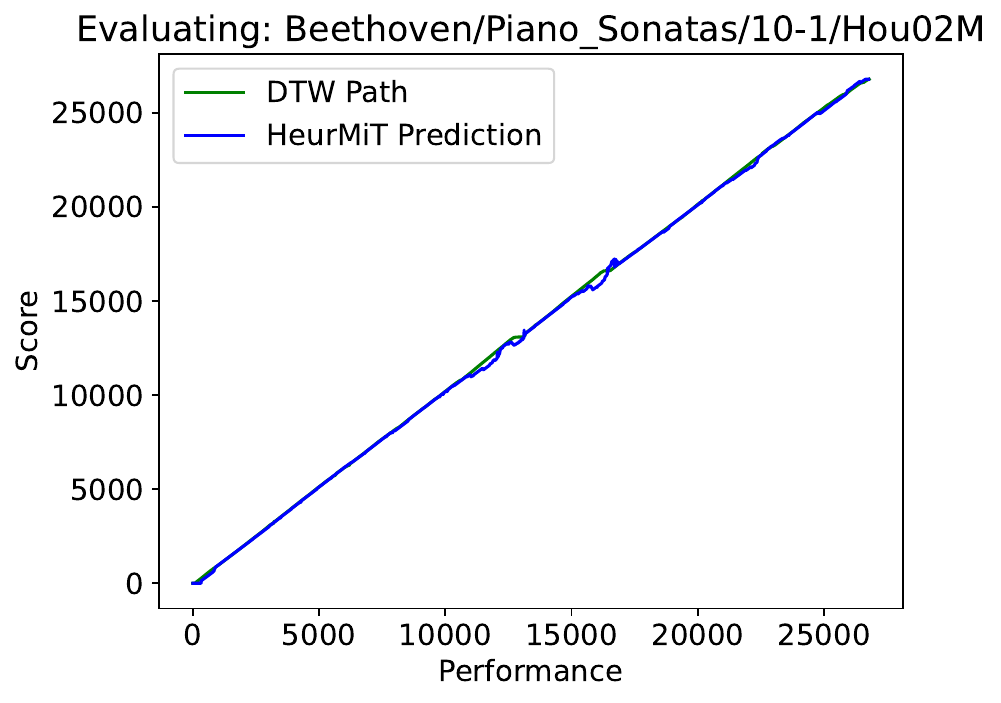}
         \captionof{figure}{P18}
         \label{fig:figure3}
      \end{minipage} &
      \begin{minipage}{.5\textwidth}
         \centering
         \includegraphics[width=.9\linewidth]{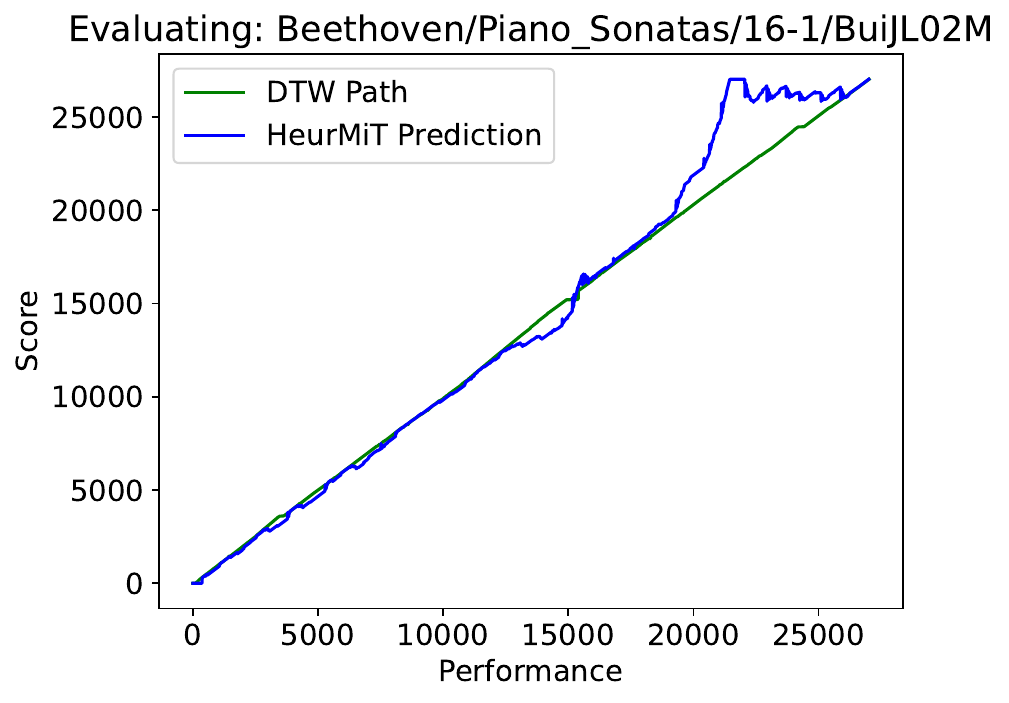}
         \captionof{figure}{P20}
         \label{fig:figure4}
      \end{minipage} \\

      \begin{minipage}{.5\textwidth}
         \centering
         \includegraphics[width=.9\linewidth]{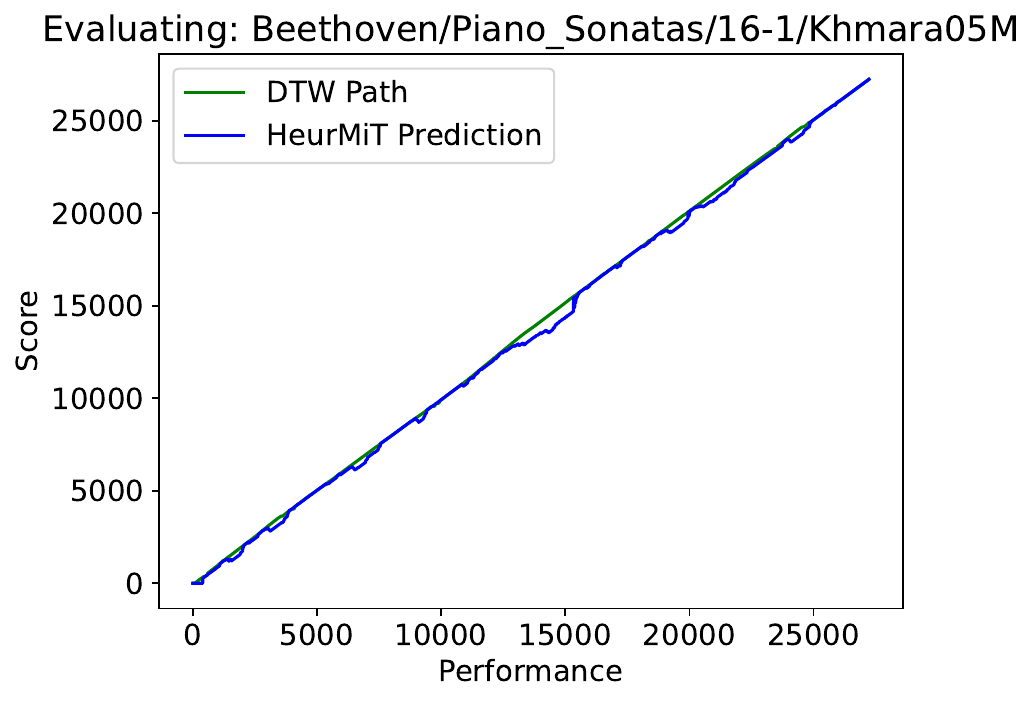}
         \captionof{figure}{P21}
         \label{fig:figure3}
      \end{minipage} &
      \begin{minipage}{.5\textwidth}
         \centering
         \includegraphics[width=.9\linewidth]{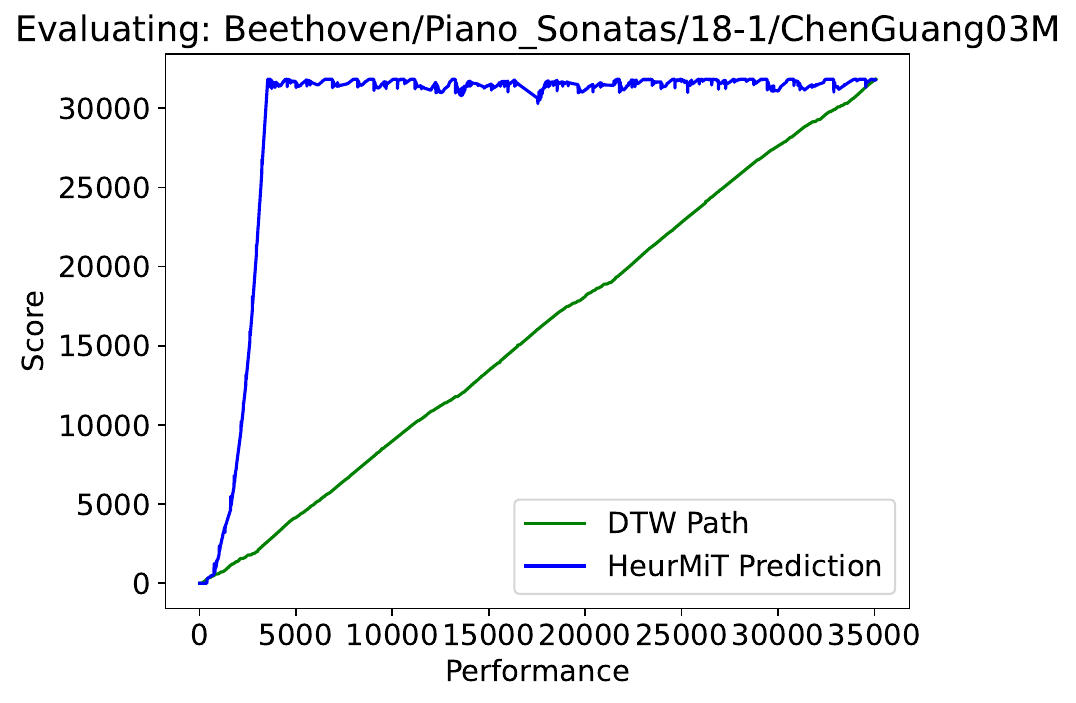}
         \captionof{figure}{P25}
         \label{fig:figure4}
      \end{minipage} \\
   \end{tabular}
\end{table}

\begin{table}[ht]
   \centering
   \begin{tabular}{cc}

      \begin{minipage}{.5\textwidth}
         \centering
         \includegraphics[width=.9\linewidth]{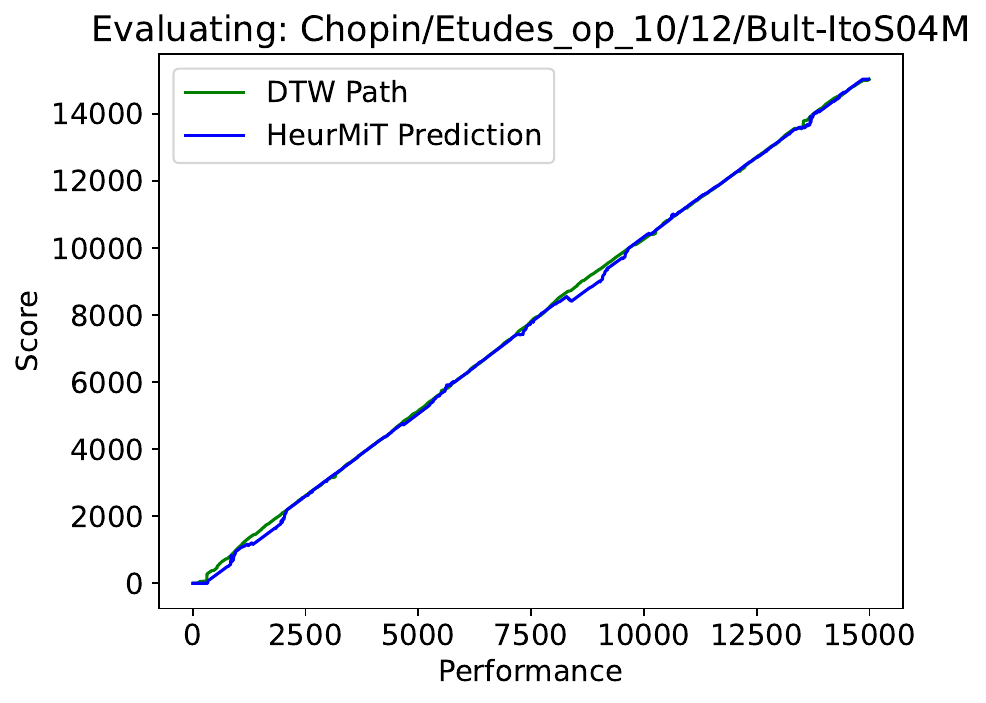}
         \captionof{figure}{P34}
         \label{fig:figure3}
      \end{minipage} &
      \begin{minipage}{.5\textwidth}
         \centering
         \includegraphics[width=.9\linewidth]{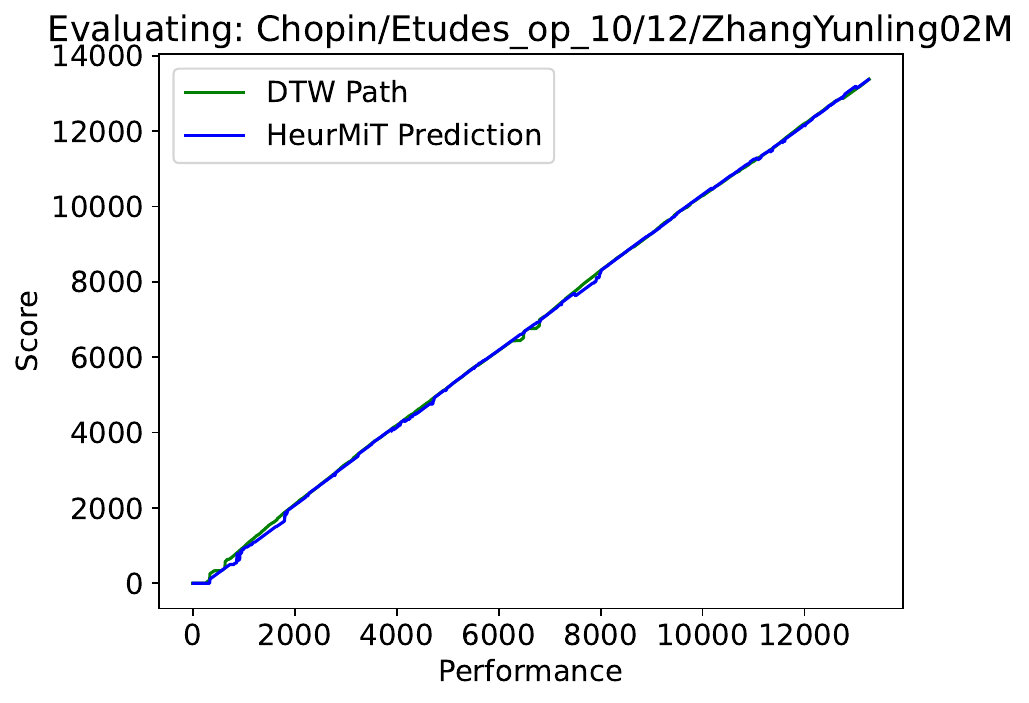}
         \captionof{figure}{P38}
         \label{fig:figure4}
      \end{minipage} \\

      \begin{minipage}{.5\textwidth}
         \centering
         \includegraphics[width=.9\linewidth]{assets/p38.pdf}
         \captionof{figure}{P39}
         \label{fig:figure1}
      \end{minipage} &
      \begin{minipage}{.5\textwidth}
         \centering
         \includegraphics[width=.9\linewidth]{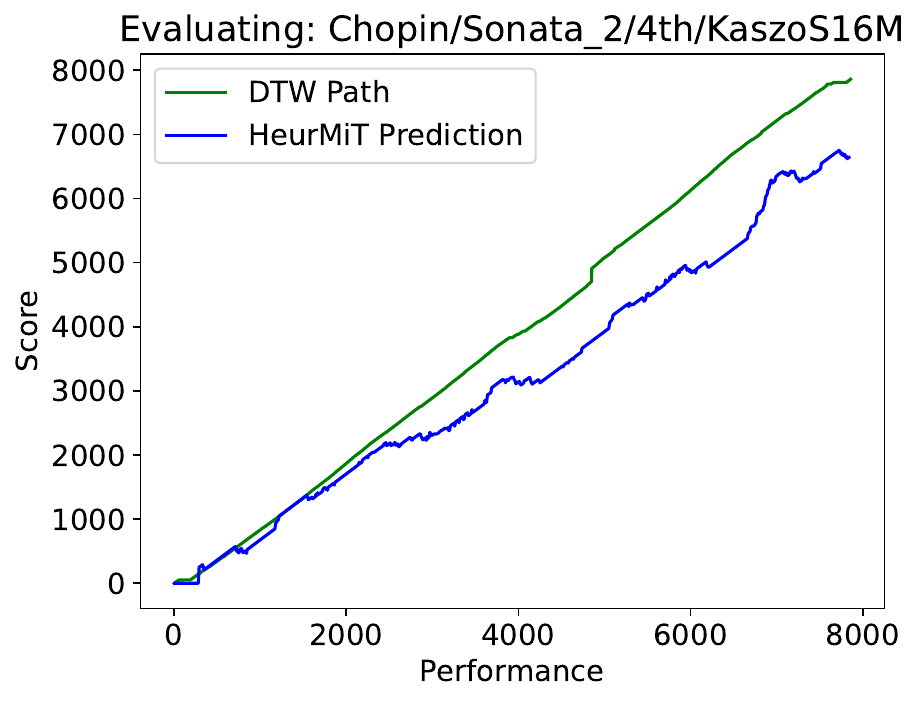}
         \captionof{figure}{P39}
         \label{fig:figure2}
      \end{minipage} \\

      \begin{minipage}{.5\textwidth}
         \centering
         \includegraphics[width=.9\linewidth]{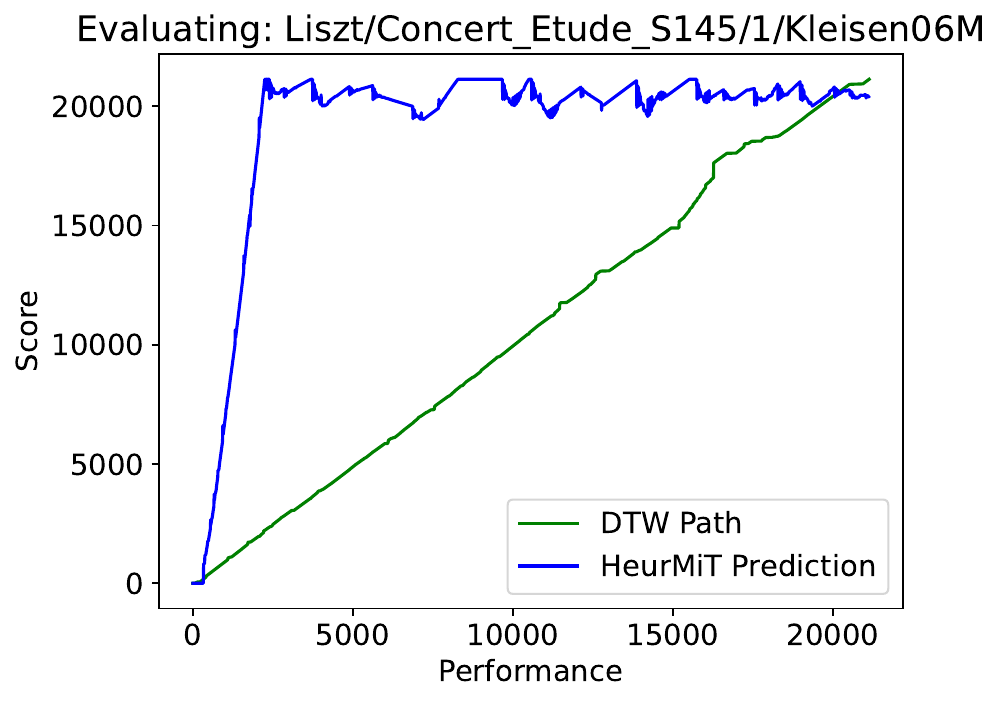}
         \captionof{figure}{P41}
         \label{fig:figure3}
      \end{minipage} &
      \begin{minipage}{.5\textwidth}
         \centering
         \includegraphics[width=.9\linewidth]{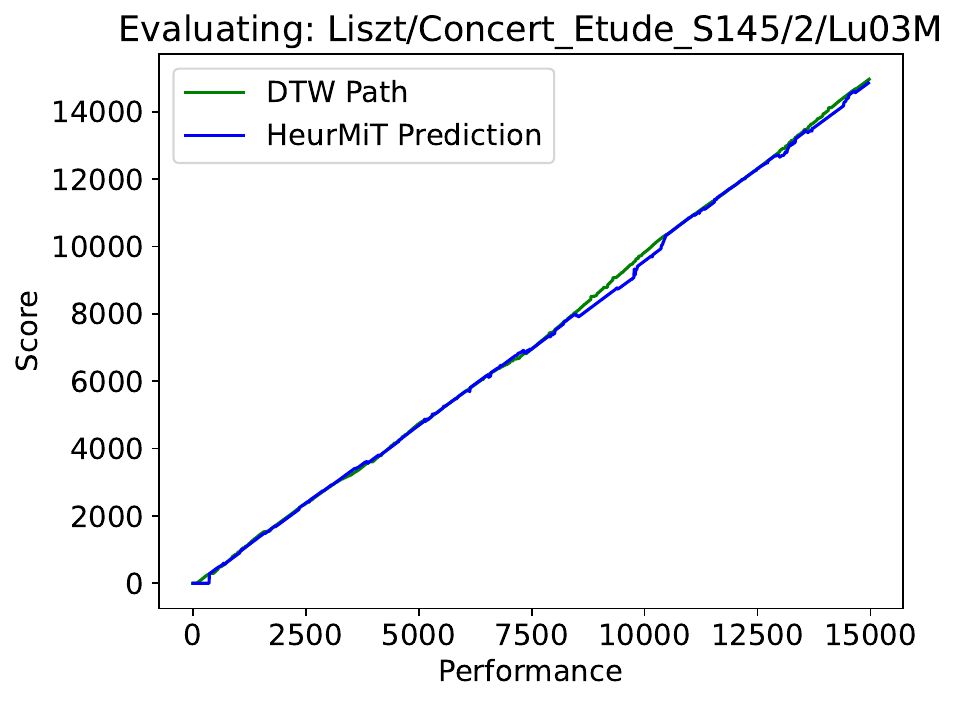}
         \captionof{figure}{P43}
         \label{fig:figure4}
      \end{minipage} \\
   \end{tabular}
\end{table}


\printbibliography[title={References},heading=bibintoc]





\end{document}